\documentclass[envcountsect,envcountsame,envcountreset]{svjour3}                     

\smartqed  
\usepackage{graphicx,amsmath,amssymb,latexsym,natbib}
\newtheorem{thm}{Theorem}[section]
\newtheorem{lem}{Lemma}[section]
\newtheorem{prop}{Proposition}[section]

\def\1{{{\mbox{${\rm{1\negthinspace\negthinspace I}}$}}}}

\begin{document}

\title{Nonparametric estimation for a stochastic volatility model.
}

\titlerunning{Nonparametric estimation for a stochastic volatility model.}        

\author{F. Comte \and V. Genon-Catalot \and Y. Rozenholc}

\authorrunning{F. Comte, V. Genon-Catalot and Y. Rozenholc} 

\institute{F. Comte \email{fabienne.comte@univ-paris5.fr} \and V.
Genon-Catalot \email{genon@math-info.univ-paris5.fr} \and Y.
Rozenholc \email{yves.rozenholc@math-info.univ-paris5.fr} \at MAP5
UMR 8145, University Paris Descartes.}

\date{Received: 2007}

\maketitle

\noindent Corresponding author: F. Comte, MAP5 UMR 8145,\\ Universit\'e Paris Descartes,\\ 45 rue des Saints-P\`eres,\\ 75270 Paris cedex 06, FRANCE.\\ email: fabienne.comte@univ-paris5.fr\\

\begin{abstract}
Consider discrete time observations $(X_{\ell\delta})_{1\leq \ell \leq n+1}$ of the process $X$ satisfying $dX_t=
\sqrt{V_t} dB_t$, with $V_t$ a one-dimensional positive diffusion process independent of the Brownian motion $B$.
For both the  drift and the diffusion coefficient  of the unobserved diffusion
$V$, we propose 
nonparametric  least   square  estimators,   and  provide  bounds   for  their
risk. Estimators are chosen 
among a collection of functions belonging to a finite dimensional space whose dimension is selected by a data driven procedure. Implementation on simulated data illustrates how the method works. \today 
\end{abstract}

\section{Introduction}\label{intro}
In this paper, we consider a bivariate process $(X_t,V_t)_{t\geq
0}$ with dynamics described by the following equations:
\begin{equation}\label{model}\left\{\begin{array}{l}
dX_t=\sqrt{V_t} dB_t, \;\; X_0=0,\\ dV_t = b(V_t) dt + \sigma(V_t)dW_t \;\;
V_0=\eta, \;\; V_t > 0, \mbox{ for all } t\geq  0,
 \end{array} \right. \end{equation}  where $(B_t,W_t)_{t\geq 0}$ is a
standard bidimensional Brownian motion and $\eta$ is independent of $(B_t,W_t)_{t\geq 0}$. Our aim is to propose and
study nonparametric estimators of $b(.)$ and $\sigma^2(.)$ on the
basis of discrete time observations of the process $X$ only.

Model (\ref{model}) was introduced by Hull and White~(1987) under the name of Stochastic Volatility model. It is often
adopted in finance to model stock prices, stock indexes or
short term interest rates: see for instance Hull and White~(1987),
Anderson and Lund~(1997), the review of Stochastic Volatility
models in Ghysels {\it et al.}~(1996) or the recent book by
Shephard~(2005) and the references therein. See also an econometric analysis of the subject in Barndorff-Nielsen and Shephard~(2002). 

The approach to study model (\ref{model}) is often parametric: the
unknown functions are specified up to a few unknown parameters,
see the popular examples of Heston~(1993) or Cox, Ingersoll and
Ross~(1985). General statistical parametric approaches of the
problem are studied in Genon-Catalot {\it et al.}~(1999), Hoffmann~(2002), Gloter~(2007), A\"{\i}t-Sahalia and Kimmel~(2007). A nonparametric estimation of the stationary density of $V_t$ is studied in Comte and Genon-Catalot~(2006). A recent proposal for nonparametric
estimation of the drift and diffusion coefficients of $V$ can be found in Ren\'o~(2006), who studies the empirical
performance of a Nadaraya-Watson kernel strategy on two parametric
simulated examples. 
Our approach is new and different, and it is based on a
nonparametric mean square strategy. We consider the same
probabilistic and sampling settings as Gloter~(2007) and follow
the ideas developed in Comte {\it et al.}~(2006, 2007), where
direct or integrated discrete observations of the process $(V_t)$
are considered. Here, our assumptions ensure that $(V_t)$  is
stationary and we consider discrete time observations
$(X_{\ell \delta})_{1\leq \ell \leq n+1}$ of the process $(X_t)$ in the
so-called high frequency context: $\delta$ is small, $n$ is large
and $n\delta=T$, the time interval where observations are taken, is large.

We assume that $n=kN$ and define as it is usual, for $i=0,1, \dots, N-1$, the realized quadratic variation associated with
$(X_{\ell \delta})_{ik+1\leq \ell < (i+1)k}$:
$$\hat{\bar V}_i=\frac 1{k\delta}\sum_{j=0}^{k-1} \left(
X_{(ik+j+1)\delta} - X_{(ik+j)\delta}\right)^2.$$
Setting $\Delta=k\delta$, $\hat{\bar V}_i$ provides an approximation of
the integrated volatility:
\begin{equation}\label{thevolint}
\bar V_i=\frac 1{\Delta}
\int_{i\Delta}^{(i+1)\Delta}V_sds,\end{equation} which in turn may
be, for well chosen $k, \delta$, a satisfactory approximation of
$V_{i\Delta}$. We have in mind to obtain regression-type equations, for $\ell=1, 2$:
$$Y_{i+1}^{(\ell)} = f^{(\ell)}(\hat{\bar V}_i) + {\rm noise}
+ {\rm remainder},$$ where \begin{equation}\label{code} f^{(1)}=b,
\; Y_i^{(1)}=\frac{\hat{\bar V}_{i+1} - \hat{\bar
V}_i}{\Delta} \mbox{ and } f^{(2)} =\sigma^2, \; Y_i^{(2)} = \frac 32 \frac{(\hat{\bar V}_{i+1} - \hat{\bar
V}_i)^2}{\Delta}.\end{equation} Choosing a collection of finite
dimensional spaces, we use the regression-type equations to
construct estimators on these spaces. Then, we propose a data
driven procedure to select a relevant estimation space in the
collection. As it is usual with these methods, the risk of an
estimator $\tilde f$ of $f=b$ or $\sigma^2$ is measured via
${\mathbb E}(\|f-\tilde f\|^2_N)$ where $\|f-\tilde f\|_N^2=(1/N)
\sum_{i=0}^{N-1} (f-\tilde f)^2(\hat{\bar V}_i)$. We obtain risk
bounds which can be interpreted as $n, N$ tend to infinity,
$\delta, \Delta$ tend to 0 and $T=n\delta=N\Delta$ tends to infinity. These bounds 
are compared with Hoffmann's~(1999) minimax rates in the case of
direct observations of $V$. For what concerns $b$, our method leads to the best rate that can be expected. 
For what concerns $\sigma^2$, no benchmark is available in this asymptotic framework. Indeed, Gloter~(2000) and 
Hoffmann~(2002) only treat the case of observations within a fixed length time interval, in a parametric setting.  As it is always the case, the rates are different for the two functions.

The paper is organized as follows. Section 2 describes the
assumptions on the model and the collection of estimation spaces.
In Section 3, the estimators are defined and their risks are
studied. Section 4 completes the procedure by the data driven
selection of the estimation space. Examples of models and
simulation results are presented in Section 5. Lastly, proofs are
gathered in Section 6.


\section{The assumptions}

\subsection{Model assumptions.}
Let $(X_t,V_t)_{t\geq 0}$ be given by (\ref{model}) and assume that
only discrete time observations of $X$,  $(X_{\ell\delta} )_{1\leq
\ell \leq n+1}$ are available. We want to
estimate the drift function $b$ and the square of the diffusion
coefficient $\sigma^2$ when  $V$ is stationary and exponentially
$\beta$-mixing. We assume that the state space of $(V_t)$ is a
known open interval $(r_0,r_1)$ of ${\mathbb R}^+$ and  consider the
following set of assumptions.

\begin{enumerate}
\item[[A1]]  $0 \leq r_0<r_1\leq +\infty$, $\stackrel{\; \; \circ}{I}=(r_0, r_1)$,  with $\sigma(v)>0$, for all $v\in\stackrel{\; \; \circ}{I}$. Let $I=[r_0,r_1]\cap {\mathbb R}$.
The function $b$ belongs to  $C^1(I)$, $b'$ is bounded on $I$, $\sigma^2\in C^2(I)$, $(\sigma^2)'\sigma$ is Lipschitz on I, $(\sigma^2)''$ is bounded on $I$
and $\sigma^2(v)\leq \sigma_1^2$ for all $v$ in $I$.
\item[[A2]] For all $v_0, v\in \stackrel{\; \; \circ}{I}$, the scale density
$s(v)=\exp \left[-2\int_{v_0}^v b(u)/\sigma^2(u)du\right]$
satisfies $\int_{r_0} s(x)dx=+\infty=\int^{r_1}
s(x)dx$, and the speed density $m(v)=1/(\sigma^2(v)s(v))$ satisfies
$\int_{r_0}^{r_1} m(v)dv=M<+\infty$.
\item[[A3]] $\eta\sim \pi$ and $\forall i, {\mathbb E}(\eta^{2i})<\infty$, where $\pi(v)dv = (m(v)/M)\1_{(r_0,r_1)}(v)dv$.
\item[[A4]] The process $(V_t)$ is exponentially $\beta$-mixing, {\em
    i.e.}, there exist constants $K>0, \theta>0$, such that, for all
  $t \ge 0$, $\beta_V(t)\leq Ke^{-\theta t}$.
\end{enumerate}

Under [A1]-[A3], $(V_t)$ is strictly stationary with marginal
distribution $\pi$, ergodic and $\beta$-mixing, {\em i.e.}
$\lim_{t\rightarrow +\infty} \beta_V(t) =0$. Here, $\beta_V(t)$
denotes the $\beta$-mixing coefficient of $(V_t)$ and is given by
$$\beta_V(t) = \int_{r_0}^{r_1} \pi(v)dv
\|P_t(v,dv')-\pi(v')dv'\|_{TV}.$$ The norm $\|.\|_{TV}$ is the
total variation norm and $P_t$ denotes the transition probability
of $(V_t)$ (see Genon-Catalot {\em et al.}~(2000)).  To prove our
main result, we need the stronger mixing condition [A4], which is
satisfied in most standard examples. Under [A1]-[A4], for fixed
$\Delta$, $(\bar V_i)_{i\geq 0}$ is a strictly stationary
process. And we have:
\begin{prop}\label{beta}
Under {\rm [A1]-[A4]}, for fixed $k$ and $\delta$, $(\hat{\bar V}_i)_{i\geq 0}$ is strictly stationary and $\beta_{\hat{\bar V}}(i) \leq c \beta_V(i\Delta)$ for all $i\geq 1$.
\end{prop}

\subsection{Spaces of approximation}\label{spaces}

\noindent The functions $b$ and $\sigma^2$ are estimated only on a
compact subset $A$ of the state space $\stackrel{\; \; \circ}{I}$. For
simplicity and without loss of generality, we assume from now on
that
\begin{equation}\label{interval} A=[0,1],
\mbox{ and we set } \;\; b_{A}= b 1_A, \quad \sigma_A= \sigma 1_A.
\end{equation}

To estimate $f=b, \sigma^2$, we consider a family $S_m, m\in {\mathcal M}_n$ of finite dimensional subspaces of  ${\mathbb L}_2([0,1])$ and compute a collection of estimators $\hat f_m$ where for all $m$, $\hat f_m$ belongs to 
$S_m$. Afterwards, a data driven procedure chooses among the collection of estimators the final estimator $\hat f_{\hat m}$.

We consider here simple projection spaces, namely trigonometric
spaces, $S_m, m\in {\mathcal M}_n$. The space $S_m$ is linearly spanned in ${\mathbb L}_2([0,1])$
by $\varphi_1, \dots, \varphi_{2m+1}$ with $\varphi_1(x)=1_{[0,1]}(x)$,
$\varphi_j(x)=\sqrt{2} \cos(2\pi jx)1_{[0,1]}(x)$ for even $j$'s
and $\varphi_j(x)=\sqrt{2} \sin(2\pi jx)1_{[0,1]}(x)$ for odd
$j$'s larger than 1. We have $D_m=2m+1={\rm dim}(S_m) \leq {\mathcal
D}_n$ and ${\mathcal M}_n=\{ 1, 3, \dots, {\mathcal D}_n\}$. The largest space in the collection has
maximal dimension ${\mathcal D}_n$, which is subject to constraints appearing later.

Actually, the theory requires smooth bases and regular wavelet bases
would also be adequate. 

In connection with the collection of spaces $S_m$, we need an additional assumption on the marginal density of the stationary process  $(\hat{\bar V}_i)_{i\geq 0}$:
\begin{itemize}\item[[A5]] The process $(\hat{\bar V}_i)_{i\geq 0}$ admits 
a stationary density $\pi^*$ and  there exist two positive constants $\pi_0^*$
and $\pi_1^*$ 
(independent of $n, \delta$) such that $\forall m\in {\mathcal M}_n$, $\forall
t\in S_m$, 
\begin{equation}\label{bornepi} \pi_0^*\|t\|^2\leq {\mathbb E}(t^2(\hat{\bar V}_0))\leq \pi_1^* \|t\|^2.
\end{equation}
\end{itemize}

The existence of the density $\pi^*$ is easy to obtain. The checking of  (\ref{bornepi}) is more technical. See the discussion on [A5] in Section \ref{discu}.
Below, we use the notations:
\begin{equation} \label{norms}
 \|t\|_{\pi^*}^2=\int t^2(x)\pi^*(x)dx, \;\;\; \|t\|^2 =
\int_0^1 t^2(x)dx \;\; \mbox{ and } \;\; \|t\|_{\infty}=\sup_{x\in [0,1]} |t(x)|.
\end{equation}

\section{Mean squares estimators of the drift and volatility}
\subsection{Regression equations}
 Reminding of (\ref{code}), we first prove the developments, for $\ell=1,2$:
\begin{equation}\label{decomp}
Y_{i+1}^{(\ell)} = f^{(\ell)}(\hat{\bar V}_i)  +
Z^{(\ell)}_{i+1} +  R^{(\ell)}(i+1),
\end{equation}
where the $Z^{(\ell)}_{i}$'s are noise terms (with martingale
properties) and the $R^{(\ell)}(i)$'s are negligible residual terms
given in Section \ref{proofs}. For the noise terms, we have, for
$\ell=1$ ($f^{(1)}=b$):
$$Z_i^{(1)} = \frac 1{\Delta^2}
\int_{i\Delta}^{(i+2)\Delta} \psi_{i\Delta}(u)\sigma(V_u)dW_u+
(u_{i+1,k}-u_{i,k})/\Delta,$$ with
\begin{equation}\label{psik}
\psi_{i\Delta}(u)= (u-i\Delta)\1_{[i\Delta, (i+1)\Delta[}(u) +
[(i+2)\Delta-u]\1_{[(i+1)\Delta, (i+2)\Delta[}(u).
\end{equation}
and
$$u_{i,k}= \frac 1{\Delta}\sum_{j=0}^{k-1} \left[\left(\int_{(ik+j)\delta}^{(ik+j+1)\delta} \sqrt{V_s}dB_s\right)^2 - \int_{(ik+j)\delta}^{(ik+j+1)\delta} V_sds \right].$$
Note that $\hat{\bar V}_i = \bar V_i + u_{i,k}$. \\
On the other hand, for $\ell=2$ ($f^{(2)}=\sigma^2$), we have $Z_i^{(2)}=Z_i^{(2,1)} +
Z_i^{(2,2)} + Z_i^{(2,3)}$ with
$$
Z_i^{(2,1)}=  \frac 3{2\Delta^3} \left[
\left(\int_{i\Delta}^{(i+2)\Delta}\psi_{i\Delta}(s)\sigma(V_s)
dW_s\right)^2 -\int_{i\Delta}^{(i+2)\Delta}
\psi_{i\Delta}^2(s)\sigma^2(V_s)ds\right],
$$
\begin{eqnarray*}
Z^{(2,2)}_i&=&\frac 3{\Delta} b(V_{i\Delta})
\int_{i\Delta}^{(i+2)\Delta}\psi_{i\Delta}(s) \sigma(V_s)dW_s \\
&& \hspace{2cm} +  \frac 3{\Delta^3} \int_{i\Delta}^{(i+2)\Delta}
\left(\int_s^{(i+2)\Delta}\psi^2_{i\Delta}(u)du\right)
[(\sigma^2)'\sigma](V_s)dW_s,
\end{eqnarray*}
where $\psi_{i\Delta}$ is given in (\ref{psik}), and
$$Z^{(2,3)}_i=\frac 3{\Delta} (\bar V_{i+1}-\bar V_i)(u_{i+1, k}-u_{i,k}).$$

\subsection{Mean squares contrast}

Equation (\ref{decomp}) gives a natural regression equation to
estimate $f^{(\ell)}$. In light of this, we consider the following
contrast, for a function $t\in S_m$ where $S_m$ is a space of the
collection and for $\ell=1,2$:
\begin{equation}\label{contrastb}
\gamma_N^{(\ell)}(t)= \frac 1N\sum_{i=0}^{N-1} [Y_{i+1}^{(\ell)}-t(\hat{\bar V}_i )]^2.\end{equation} Then the
estimators are defined as
\begin{equation}\label{defdeb}
\hat f_m^{(\ell)} =\arg\min_{t\in S_m} \gamma_N^{(\ell)}(t).
\end{equation}
The minimization of  $\gamma_N^{(\ell)}$ over $S_m$ usually leads to several solutions.
In contrast, the random ${\mathbb
R}^N$-vector $(\hat f_m^{(\ell)}(\hat{\bar V}_0), \dots, \hat
f_m^{(\ell)}(\hat{\bar V}_{N-1}))'$ is always uniquely defined.
Indeed, let us denote by $\Pi_m$ the orthogonal projection (with
respect to the inner product of ${\mathbb R}^N$) onto the subspace
of ${\mathbb R}^N$, $\{(t(\hat{\bar V}_0), \dots, t(\hat{\bar
V}_{N-1}))', t\in S_m\}$, then $(\hat f^{(\ell)}_m(\hat{\bar V}_0),
\dots, f^{(\ell)}_m(\hat{\bar V}_{N-1}))'=\Pi_m  Y^{(\ell)}$
where  $Y^{(\ell)}=(\bar Y_1^{(\ell)}, \dots,  Y_{N}^{(\ell)})'$. This  is the
reason why we consider a 
properly defined risk for $\hat f_m^{(\ell)}$ based on the design points, i.e.
$${\mathbb E}\left[\frac 1N\sum_{i=0}^{N-1} (\hat f_m^{(\ell)}(\hat{\bar V}_i)-f(\hat{\bar V}_i))^2\right].$$
Thus, the error is measured via the risk ${\mathbb E}(\|\hat
f_m^{(\ell)}-f^{(\ell)}\|^2_N)$ where $$\|t\|_N^2=\frac 1N\sum_{i=0}^{N-1}
t^2(\hat{\bar V}_i).$$
Let us mention that for a deterministic function
${\mathbb E}(\|t\|^2_N)=\|t\|_{\pi^*}^2=\int t^2(x)\pi^*(x)dx$. Moreover, under Assumption [A5],
the norms $\|.\|$ and $\|.\|_{\pi^*}$ are equivalent for functions in $S_m$ (see notations (\ref{norms})).\\

The following decomposition of the contrast holds:
\begin{eqnarray*} \gamma_N^{(\ell)}(t)-\gamma_N^{(\ell)}(f^{(\ell)})&=&\|t-f^{(\ell)}\|_N^2
- \frac 2{N}\sum_{i=0}^{N-1}(Y_{i+1}^{(\ell)} -f^{(\ell)}(\hat{\bar V}_i)) (f^{(\ell)}-t)(\hat{\bar V}_i)
\end{eqnarray*}
In view of (\ref{decomp}), we define the
centered empirical processes, for $\ell=1,2$:
$$\nu_N^{(\ell)}(t)= \frac 1{N}\sum_{i=0}^{N-1} t(\hat{\bar V}_i^{(\ell)})Z^{(\ell)}_{i+1},$$
and the residual process:
$$R_N^{(\ell)}(t)=\frac 1{N}\sum_{i=0}^{N-1} t(\hat{\bar V}_i) R^{(\ell
)}(i+1).$$
Then we obtain that
\begin{eqnarray*} \gamma_N^{(\ell)}(t)-\gamma_N^{(\ell)}(f^{(\ell)})&=&\|t-f^{(\ell)}\|_N^2 -
2\nu_N^{(\ell)}(t-f^{(\ell)})- 2R_N^{(\ell)}(t-f^{(\ell)}).
\end{eqnarray*}

Let $f_m^{(\ell)}$ be the orthogonal projection of $f^{(\ell)}$ on
$S_m$. Write simply that $\gamma_N^{(\ell)}(\hat f_m^{(\ell)})\leq
\gamma_N^{(\ell)}(f_m^{(\ell)})$ by definition of the estimator,
and therefore that $\gamma_N^{(\ell)}(\hat
f_m^{(\ell)})-\gamma_N^{(\ell)}(f^{(\ell)})\leq
\gamma_N^{(\ell)}(f^{(\ell)}_m)-\gamma_N^{(\ell)}(f^{(\ell)})$. This yields  
\begin{eqnarray*} \|\hat f_m^{(\ell)}-f^{(\ell)}\|_N^2&\leq & \|f^{(\ell)}_m-f^{(\ell)}\|_N^2 + 2\nu_N^{(\ell)}(\hat f^{(\ell)}_m-f^{(\ell)}_m) + 2R_N^{(\ell)}(\hat f^{(\ell)}_m-f^{(\ell)}_m).
\end{eqnarray*}
The functions $\hat f_m^{(\ell)}$ and $f_m^{(\ell)}$ being
$A$-supported, we can cancel the terms $\|f\1_{A^c}\|^2_N$ that
appears in both sides of the inequality. Therefore, we get 
\begin{equation}\label{deccontrast}  \|\hat f_m^{(\ell)}-f_A^{(\ell)}\|_N^2 \leq  \|f^{(\ell)}_m-f_A^{(\ell)}\|_N^2 + 2\nu_N^{(\ell)}(\hat f^{(\ell)}_m-f^{(\ell)}_m) + 2R_N^{(\ell)}(\hat f^{(\ell)}_m-f^{(\ell)}_m).
\end{equation}
Taking expectations and finding upper bounds for  $${\mathbb E}(\sup_{t\in S_m, \|t\|=1}[\nu_N^{(\ell)}(t)]^2) \; \; 
\mbox{ and  } \;\; {\mathbb E}(\sup_{t\in S_m, \|t\|=1}[R_N^{(\ell)}(t)^2)$$ will give the rates for the risks of the estimators.

\subsection{Risk for the collection of drift estimators}
For the estimation of $b$, we obtain the following result.
\begin{prop}\label{propb} Assume that $N\Delta\geq 1$ and
$1/k\leq \Delta$. Assume that {\rm [A1]-[A5]} hold and consider a
model $S_m$ in the collection of models with ${\mathcal D}_n\leq
O(\sqrt{N\Delta}/\ln(N))$ where ${\mathcal D}_n$ is the maximal dimension (see Section \ref{spaces}).
Then the estimator $\hat f_m^{(1)}=\hat b_m$
of $f^{(1)}=b$ is such that
\begin{equation}\label{borneb}
{\mathbb E}(\|\hat b_m-b_A\|_n^2) \leq  7 \|b_m-b_A\|^2_{\pi^*} +
 K\frac{{\mathbb E}(\sigma^2(V_0)) D_m}{N\Delta}
 + K'\Delta,
\end{equation}
where $b_A=b\1_{[0,1]}$ and $K, K'$ and $K"$ are some positive
constants.
\end{prop}
Note that the condition on ${\mathcal D}_n$ implies that
$\sqrt{N\Delta}/\ln(N)$ must be large enough.

It follows from (\ref{borneb}) that it is natural to
select the dimension $D_m$ that leads to the best
compromise between the squared bias term $\|b_m-b_A\|^2_{\pi^*}$
(which decreases when $D_m$ increases) and the variance term of
order $D_m/(N\Delta)$. \\

Now, let us consider the classical high frequency data setting:
let $\Delta=\Delta_n$, $k=k_n$ and $N=N_n$ be, in addition, such
that $\Delta_n \rightarrow 0$, $N=N_n\rightarrow +\infty$, $\;N_n\Delta_n/\ln^2(N_n) \rightarrow +\infty$ when $n\rightarrow +\infty$ and that $1/(k_n\Delta_n)\leq 1$. Assume for instance that $b_A$ belongs to a ball of some
Besov space, $b_A\in {\mathcal B}_{\alpha,2,\infty}([0,1])$,
$\alpha\geq 1$, and that $\|b_m-b_A\|^2_{\pi^*}\leq \pi_1^*
\|b_m-b_A\|^2$, then $\|b_A-b_m\|^2_{\pi^*} \leq
C(\alpha,L,\pi_1^*) D_m^{-2\alpha}$, for $\|b_A\|_{\alpha,
2,\infty}\leq L$ (see Lemma 12 in Barron {\it et al.}~(1999)). Therefore, if we choose
$D_m=(N_n\Delta_n)^{1/(2\alpha+1)}$,
we obtain
\begin{equation}\label{vitesb} {\mathbb E}(\|\hat b_m-b_A\|_n^2) \leq  C(\alpha, L) (N_n\Delta_n)^{-2\alpha/(2\alpha+1)} + K'\Delta_n.\end{equation}
The first term $(N_n\Delta_n)^{-2\alpha/(2\alpha +1)}=T_n^{-2\alpha/(2\alpha +1)}$ is the
optimal nonparametric rate proved by Hoffmann~(1999) for direct
observation of $V$.

Now, let us find conditions under which the last term is 
negligible. For instance, under the standard condition
$\Delta_n=O(1/(N_n\Delta_n))$, the term $\Delta_n$ is negligible with respect to 
$(N_n\Delta_n)^{-2\alpha/(2\alpha +1)}$.

Now, consider the choices  $k_n=1/\Delta_n$ and $\delta_n=n^{-c}$. Let us see if there are possible choices of $c$ for which all our constraints are fulfilled.  
To have $n\delta_n\rightarrow +\infty$ requires $0<c<1$. As $\Delta_n=k_n\delta_n=\delta_n/\Delta_n$, we have
$\Delta_n=\sqrt{\delta_n}=n^{-c/2}$ and $N_n=n/k_n=n^{1-c/2}$. Thus, $\Delta_n\rightarrow 0$ and  $N_n,
N_n\Delta_n \rightarrow +\infty$. Finally, the last constraint to fulfill is that $N_n\Delta_n^2=n^{1-3c/2}=O(1)$. Thus for $2/3\leq c<1$, the dominating term in (\ref{vitesb}) is $(N_n\Delta_n)^{-2\alpha/(2\alpha +1)}$, {\it i.e.} the minimax optimal rate. We have obtained a possible ``bandwidth" of steps $\delta_n$.


\subsection{Risk for the collection of volatility estimators}
For the collection of volatility estimators, we have the result
\begin{prop}\label{bornesigma}
 Assume that {\rm [A1]-[A5]} hold and
consider a model $S_m$ in the collection of models with maximal dimension ${\mathcal
D}_n\leq O(\sqrt{N\Delta}/\ln(N))$. Assume also that $1/k\leq \Delta$
and $N\Delta\geq 1$, $\Delta\leq 1$. Then the estimator $\hat
f_m^{(2)}=\hat \sigma^2_m$ of $f^{(2)}=\sigma^2$
 is such that
\begin{equation}\label{bornes}
{\mathbb E}(\|\hat \sigma^2_m-\sigma_A^2\|_N^2) \leq   7
\|\sigma^2_m-\sigma_A^2\|_{\pi^*}^2 + K\frac{{\mathbb E}(\sigma^4(V_0)) D_m}{N} + K'Res(D_m, k, \Delta),
\end{equation}
where the residual term is given by 
\begin{equation}\label{shortres}
Res(D_m,k,\Delta) = D_m^2\Delta^2 + D_m^5\Delta^3+  \frac{D_m^3}{k^2} +
 \frac 1{k^2\Delta^2},
\end{equation}
where $\sigma^2_A=\sigma^2\1_{[0,1]}$, and $K$, $K'$ are some positive constants.
\end{prop}

The discussion on rates is much more tedious. 
Consider the asymptotic setting described for $b$. Assume that $\sigma^2_A$ belongs to a ball of some Besov space,
$\sigma^2_A\in {\mathcal B}_{\alpha,2,\infty}([0,1])$, and that
$\|\sigma^2_m-\sigma_A^2\|_{\pi^*}^2 \leq \pi_1^*
\|\sigma^2_m-\sigma_A^2\|^2$, then
$\|\sigma^2_A-\sigma^2_m\|^2_{\pi^*}\leq C(\alpha,L ,\pi_1^*)
D_m^{-2\alpha}$, for $\|\sigma^2_A\|_{\alpha, 2,\infty}\leq L$. Therefore, if we choose $D_m=N_n^{1/(2\alpha+1)}$, and $k_n \leq
1/\Delta_n$, we obtain
\begin{equation}\label{vitessig} {\mathbb E}(\|\hat \sigma^2_m-\sigma^2_A\|_N^2) \leq
C(\alpha, L, \pi_1^*) N_n^{-2\alpha/(2\alpha+1)} + K' Res(N_n^{1/(2\alpha+1)}, k_n, \Delta_n).
\end{equation} The first term
$N_n^{-2\alpha/(2\alpha +1)}$ is the optimal nonparametric rate
proved by Hoffmann~(1999) when $N_n$ discrete time observations of
$V$ are available.

For the second term, let us set $k_n=n^a$, $\Delta_n=n^{-b}$, $\delta_n=n^{-c}$, and recall that $n\delta_n=N_n\Delta_n$ and $n/N_n=k_n$, so that $N_n=n^{1-a}$ and $a+b=c$. We look for $a, b$ such that
$$Res(N_n^{1/(2\alpha+1)}, k_n, \Delta_n) \leq N_n^{-2\alpha/(2\alpha+1)}.$$
For this, we take $1/(k_n^2\Delta_n^2)=N_n^{-2\alpha/(2\alpha+1)}$ which implies 
$2(a-b)/(1-a)=2\alpha/(2\alpha+1)$. We get 
$$a=\frac{(2\alpha+1)c+\alpha}{5\alpha+2}, \;\;
b=\frac{(3\alpha+1)c-\alpha}{5\alpha+2}.$$ Then we impose
$N_n^{2/(2\alpha+1)}\Delta_n^2\leq N_n^{-2\alpha/(2\alpha+1)}$ which is equivalent to $$2b\geq [(2\alpha+2)/(2\alpha+1)](1-a) \Rightarrow c\geq (3\alpha+2)[2(2\alpha+1)].$$
Next
$N_n^{5/(2\alpha+1)}\Delta_n^3 \leq N_n^{-2\alpha/(2\alpha+1)}$ leads to 
$$3b\geq [(2\alpha+5)/(2\alpha+1)](1-a) \Rightarrow c\geq (7\alpha+5)/(11\alpha+8).$$ Lastly
$N_n^{3/(2\alpha+1)}/k_n^2\leq N_n^{-2\alpha/(2\alpha+1)}$ holds for $-2a\leq -[(3+2\alpha)/(2\alpha +1)](1-a)$, {\em i.e.} $c\geq 2(\alpha+3)/(6\alpha+5)$.

The optimal dimension has also to fulfill $N_n^{1/(2\alpha+1)}\leq {\mathcal D}_n\leq 
\sqrt{N_n\Delta_n}$ {\it i.e.}  $-[(2\alpha-1)/[2(2\alpha+1)]](1-a)\leq -b/2$ which
implies $c\leq (5\alpha -2)/(5\alpha)$. Finally, we must have
$$c\in \left[\frac{3\alpha+2}{2(2\alpha+1)},
\frac{5\alpha-2}{5\alpha}\right] \rightarrow_{\alpha\rightarrow
+\infty} \left] \frac 34, 1\right[.$$ This interval is nonempty as soon as $\alpha>2$.

In terms of the initial number $n$ of observations, the rate is
now $(n^{1-a})^{-2\alpha/(2\alpha+1)}$ where $1-a$ is at most
$1/2$, when $\alpha \rightarrow+\infty$. This is consistent with Gloter's~(2000) result: in the
parametric case, he obtains $n^{-1/2}$ instead of $n^{-1}$ for the quadratic risk.


\section{Data driven estimator of the coefficients}

The second step is to ensure an automatic selection of $D_m$,
which does not use any knowledge on $f^{(\ell)}$, and in
particular which does not require to know the regularity $\alpha$. This selection
is standardly done by setting \begin{equation}\label{selecm} \hat
m^{(\ell)}=\arg\min_{m\in {\mathcal M}_n} \left[
\gamma_n^{(\ell)}(\hat f_m^{(\ell)}) + {\rm pen}^{(\ell)}(m)
\right],\end{equation} with pen$^{(\ell)}(m)$ a penalty to be
properly chosen. We denote by $\tilde f^{(\ell)}=\hat
f^{(\ell)}_{\hat m^{(\ell)}}$ the resulting estimator and we need
to determine pen such that, ideally,
$${\mathbb E}(\|\tilde f^{(\ell)}-f_A^{(\ell)}\|_n^2) \leq
C\inf_{m\in {\mathcal M}_n}\left( \|f^{(\ell)}_A-f_m^{(\ell)}\|^2
+ \frac{{\mathbb E}(\sigma^{2\ell}(V_0)) D_m}{N\Delta^{2-\ell}}\right) +
\mbox{ negligible terms,} $$ with $C$ a constant which should not be too
large.

\subsection{Result for the data driven estimator of $b$}
We almost reach this aim for the estimation of $b$.
\begin{thm}\label{maindrift}
Assume that {\rm [A1]-[A5]} hold, $1/k\leq \Delta$, $\Delta\leq 1$
and $N\Delta\geq 1$. Consider the collection of models with
maximal dimension ${\mathcal D}_n\leq O(\sqrt{N\Delta}/\ln(N))$. Then
the estimator $\tilde b=\hat f^{(1)}_{\hat m^{(1)}}$ of $b$ where
$\hat m^{(1)}$ is defined by (\ref{selecm}) with
\begin{equation}\label{penb} {\rm pen}^{(1)}(m)\geq  \kappa \sigma_1^2
\frac{D_m}{N\Delta},
\end{equation}
where $\kappa$ is a universal constant, is such that
\begin{eqnarray}\nonumber 
{\mathbb E}(\|\tilde b-b_A\|_n^2)& \leq & C\inf_{m\in {\mathcal
M}_n}\left( \|b_m-b_A\|^2_{\pi^*} +  {\rm pen}^{(1)}(m)\right)\\ \label{bornebadapt}
&& \hspace{2cm}  + K\left(\Delta + \frac 1{N\Delta}+
 \frac 1{\ln^2(N) k\Delta}\right).
\end{eqnarray}
\end{thm}

For comments on the practical calibration of the penalty, see Section \ref{simures}.

It follows from (\ref{bornebadapt}) that the adaptive estimator
automatically realizes the bias-variance compromise, provided that
the last terms can be neglected as discussed above. Here, the
bandwidth for the choices of $\delta_n$ is slightly narrowed
because of a stronger constraint. More precisely, we choose
$1/(k_n\Delta_n)=\Delta_n$ (instead of 1 previously), that is
$k_n=\Delta_n^{-2}$, so that
$\Delta_n=k_n\delta_n=\Delta_n^{-2}\delta_n^{-1}$. Therefore
$\Delta_n=\delta_n^{1/3}$ and if $\delta_n=n^{-c}$, then
$\Delta_n=n^{-c/3}$. Also, $N_n=n/k_n=n^{1-2c/3}$,
$N_n\Delta_n=n\delta_n=n^{1-c}$, $N_n\Delta_n^2= n^{1-4c/3}$.  Hence if $3/4<c<1$,
we have altogether: $N_n$, $N_n\Delta_n/\ln^2(N_n)$ tend to infinity with $n$, $\Delta_n$,
$N_n\Delta_n^2$ tend to zero.

In that case, whenever $b_A$ belongs to some Besov ball (see
(\ref{vitesb})), and if $\|b_m-b_A\|^2_{\pi^*}\leq \pi_1^*
\|b_m-b_A\|^2$, then $\tilde b$ achieves the optimal corresponding
nonparametric rate. Note that, in the parametric framework, Gloter~(2007) obtains an efficient estimation of $b$ in the same asymptotic context. 

\subsection{Result for the data driven estimator of the volatility}

We can prove the following Theorem.

\begin{thm}\label{mainvol}
Assume that {\rm [A1]-[A5]} hold, $1/k\leq \Delta$, $\Delta\leq 1$
and $N\Delta\geq 1$. Consider the collection of models with
maximal dimension ${\mathcal D}_n\leq \sqrt{N\Delta}/\ln(N)$. Then
the estimator $\tilde \sigma^2=\hat f^{(2)}_{\hat m^{(2)}}$ of
$\sigma^2$ where $\hat m^{(2)}$ is defined by (\ref{selecm}) with
\begin{equation}\label{pensig} {\rm pen}^{(2)}(m)\geq  \kappa
\sigma_1^4  \frac{D_m}{N},
\end{equation}
where $\kappa$ is a universal constant, is such that
\begin{equation}\label{bornesigadapt}
{\mathbb E}(\|\tilde \sigma^2-\sigma_A^2\|_N^2) \leq 
C\inf_{m\in {\mathcal M}_n}\left(\|\sigma^2_m-\sigma_A^2\|^2_{\pi^*} + {\rm pen}^{(2)}(m) \right)
+ C'\widetilde{Res}(N, k, \Delta),
\end{equation} where 
\begin{equation}\label{restilde} 
\widetilde{Res}(N, k, \Delta)
= N\Delta^3+ N^{5/2}\Delta^{11/2}+\frac{(N\Delta)^{3/2}}{k^2}+
\frac 1{k^2\Delta^2} .
\end{equation}
\end{thm}

Now, if $\sigma^2_A$ belongs to a ball of some
Besov space, $\sigma^2_A\in {\mathcal
B}_{\alpha,2,\infty}([0,1])$, then automatically,
$$\inf_{m\in {\mathcal M}_n}\left(\|\sigma^2_m-\sigma_A^2\|^2_{\pi^*} + {\rm pen}^{(2)}(m) \right)=O(N_n^{-2\alpha/(2\alpha+1)})$$
without requiring the knowledge of $\alpha$.
Therefore,  
$${\mathbb E}(\|\breve \sigma^2_{\breve m}-\sigma^2_A\|_N^2) \leq
C(\alpha, L) N_n^{-2\alpha/(2\alpha+1)} + C'\widetilde{Res}(N_n, k_n, \Delta_n).$$
It remains to study the residual term. 
Notice that we do not know the optimal minimax rate for estimating $\sigma^2$, under our set of assumptions on the models and on the asymptotic framework. However, Gloter~(2000) and Hoffmann~(2002), with observations within a fixed length time interval, obtain the parametric rate $n^{-1/2}$ (in variance). Taking this as a benchmark, we try to make the residual less than $O(n^{-1/2})$. Let us set $k_n=n^a$, $\Delta_n=n^{-b}$, hence $N_n=n/k_n=n^{1-a}$ and $N_n\Delta_n= n^{1-(a+b)}$. This yields that $1-a-3b, (5-5a-11b)/2, (3-7a-3b)/2, 2(b-a)$ must all be less than or  equal to $-1/2$, in association with $a+b<1$ and $N_n^{1/(2\alpha+1)}\leq \sqrt{N_n\Delta_n}$. This set of constraint is not empty (e.g. $a=9/16, b=5/16$ fits). 


\section{Examples and numerical simulation results}\label{simu}
In this section, we consider examples of diffusions and implement
the estimation algorithm on simulated data for the stochastic
volatility model $X$ given by (\ref{model}).

\subsection{Simulated paths}

We consider the processes $V_t^{(i)}$ for $i=1, \dots, 4$ specified by the couples of functions $b_i, \sigma_i^2$, $i=1, \dots,4$:
\begin{enumerate} \item $b_1(x)= x\left( - \theta \ln(x)+ \frac{1}2 c^2\right), \sigma_1^2(x) = c^2 x^2$ which corresponds to $\exp(U_t)$ for $U_t$ an Ornstein-Uhlenbeck process, $dU_t=-\theta U_t dt +c dW_t$. Whatever the chosen step, $U_t$ is exactly simulated as an autoregressive process of order 1. We took $\theta=1$ and $c=0.75$.
\item $b_2(x)=b_0(x-2)$, $\sigma_2^2(x)=\sigma^2_0(x-2)$, where $b_0(x)= -(1-x^2)\left[ c^2x +\frac{\theta}2 \ln\left(\frac{1+x}{1-x}\right) \right]$ and $\sigma_0^2(x)=c(1-x^2)$ are the diffusion coefficients of the process
${\rm th}(U_t)$ (th$(x)=(e^x-e^{-x})/(e^x+e^{-x})$, with the same parameters as for case $1)$. The process $V_t^{(2)}$ corresponds to th$(U_t)+2$ which is a positive bounded process.
\item $b_3(x)= x(b_0(\ln(x))+ \frac 12 \sigma_0^2(\ln(x)))$ and $\sigma_3^2(x)= x^2\sigma_0^2(\ln(x))$ which corresponds to the process $V_t^{(3)}=\exp({\rm th}(U_t))$.
\item $b_4(x)= dc^2/4 -\theta x, \sigma^2_4(x) =c^2 x$ which corresponds 
to the  Cox-Ingersoll-Ross process. A discrete  time sample is  obtained in an
exact way by taking the 
Euclidean norm of a $d$-dimensional Ornstein-Uhlenbeck process with parameters $-\theta/2$ and $c/2$. We took $d=9$, $\theta=0.75$ and $c=1/3$. 
\end{enumerate}

\noindent
We obtain samples of discrete observations of the processes
$(V_{\ell  \delta'}^{(j)})_{1\leq  \ell\leq  N'}$  for $j=1,  \dots,  4$  with
$\delta'=\delta/10$, $N'\delta'=T$, from
which we generate $(X_{\ell \delta}^{(j)})_{1\leq \ell \leq n}$, by using
that $$X_{\ell \delta}-X_{(\ell-1)\delta} =
\sqrt{\int_{(\ell-1)\delta}^{\ell \delta} V_sds} \;\; \varepsilon_{\ell},$$
with $(\varepsilon_{\ell})$
i.i.d. $\mathcal {N}(0,1)$ independent of $(V_s, s\ge 0)$. Approximations of the integrated
processes are computed by discrete integration (with a trapeze method).

The generated $V_{j\delta'}^{(i)}$, $i=1,\dots, 4$ samples have length $N'=5.10^6$, for 
a step $\delta'=1000/5.10^6=2.10^{-4}$, and the integrated process is computed
using 10 data, 
therefore, we obtain $n=5.10^5$ and $\delta=2.10^{-3}$, for $T=n\delta=1000$. 
Different values of $k$ are used,  but the best value, $k=250$, corresponds to
$\Delta=k\delta=0.5$ 
and $N=2000$ data for the same $T$.



\subsection{Estimation algorithms and numerical results}\label{simures}
\begin{table}\label{tab1}
\begin{tabular}{cc|c|c|c|c|c}
&& $k=150$ & $k=200$ & $k= 250$ & $k=300$ & $k=500$ \\ \hline
&&&&&&\\
$b$ & mean &   $1,70.10^{-3}$ & $1,87.10^{-3}$ & $1,95.10^{-3}$ & $2,1.10^{-3}$ & $2,91.10^{-3}$ \\
& (std)& ($5,38.10^{-4}$) &($5,06.10^{-4}$) & ($4,93.10^{-4}$) & ($4,92.10^{-4}$) & ($4,68.10^{-4}$)\\
&&&&&&\\\hline
&&&&&&\\
$\sigma^2$ & mean &  $14,8.10^{-5}$ & $6,23.10^{-5}$ & $8,77.10^{-5}$ & $15,3.10^{-5}$&$28,6.10^{-5}$\\
& (std) & ($3,26.10^{-5}$) &($2,26.10^{-5}$) & ($3,74.10^{-5}$) & ($4,0.10^{-5}$) & ($3,39.10^{-5}$)
\end{tabular}
\caption{Mean squared errors (with standard deviations in parenthesis) for the estimation of $b$ and $\sigma^2$, 100 paths of the CIR process, different values of $k$ for the quadratic variation, when using the trigonometric basis.}
\end{table}
\begin{table}\label{tab2}
\begin{tabular}{cc|c|c|c|c|c}
& Process &  $V_t^{(1)}$ [T] & $V_t^{(2)}$ [T] & $V_t^{(3)}$ [T] & $V_t^{(4)}$ [T] & $V_t^{(4)}$ [GP] \\\hline
&&&&&&\\
$b$ & mean &    $4,08.10^{-2}$ & $7,51.10^{-2}$ & $7,05.10^{-2}$ & $1,95.10^{-3}$  & $1,04.10^{-3}$ \\
& (std)&  ($6,89.10^{-3}$) & ($8,56.10^{-3}$) & ($8,12.10^{-3}$) & ($4,93.10^{-4}$) & ($2,89.10^{-4}$)\\
&&&&&&\\ \hline
&&&&&&\\
$\sigma^2$ & mean &  $1,42.10^{-1}$ & $1,89.10^{-2}$ & $8,32.10^{-2}$ & $8,77.10^{-5}$&$4,61.10^{-5}$\\
& (std)  &($3,47.10^{-2}$) & ($1,54.10^{-3}$) & ($1,61.10^{-2}$) & ($3,74.10^{-5}$) & ($3,19.10^{-5}$)
\end{tabular}
\caption{Mean squared errors (with standard deviations in parenthesis) for the estimation of $b$ and $\sigma^2$, 100 paths of the processes $V_t^{(i)}$, $i=1,\dots,4$ when using the trigonometric basis (except the last column, piecewise polynomial basis), $k=250$.}
\end{table}

We use the algorithm of Comte and Rozenholc (2004). The precise calibration of
penalties 
is difficult and done for the trigonometric basis but also for a general piecewise polynomial basis, described in detail in Comte~{\it et al}~(2006).
Additive correcting terms are involved in the penalty. Such terms avoid under-penalization and are in
accordance with the fact that the theorems provide lower bounds
for the penalty. The correcting terms are asymptotically negligible and do not affect the rate of convergence.
For the trigonometric polynomial collection (denoted by [T]), the drift penalty $(i=1)$ and the diffusion penalty $(i=2)$ are given by 
$$2\frac{\hat s_i^2}n\left(D_m+ \ln^{2.5}(D_m+1)\right), \mbox{ with } D_m \mbox{ at most } [N\Delta/\ln^{1.5}(N)].$$
For the penalty when considering general piecewise polynomial bases (denoted by [GP]), we refer the reader to Comte~{\it et al.}~(2006).
\begin{figure}[!ht]\label{fig1}
\begin{tabular}{c}
$\includegraphics[width=1.1\textwidth,height= 4cm,angle=0]{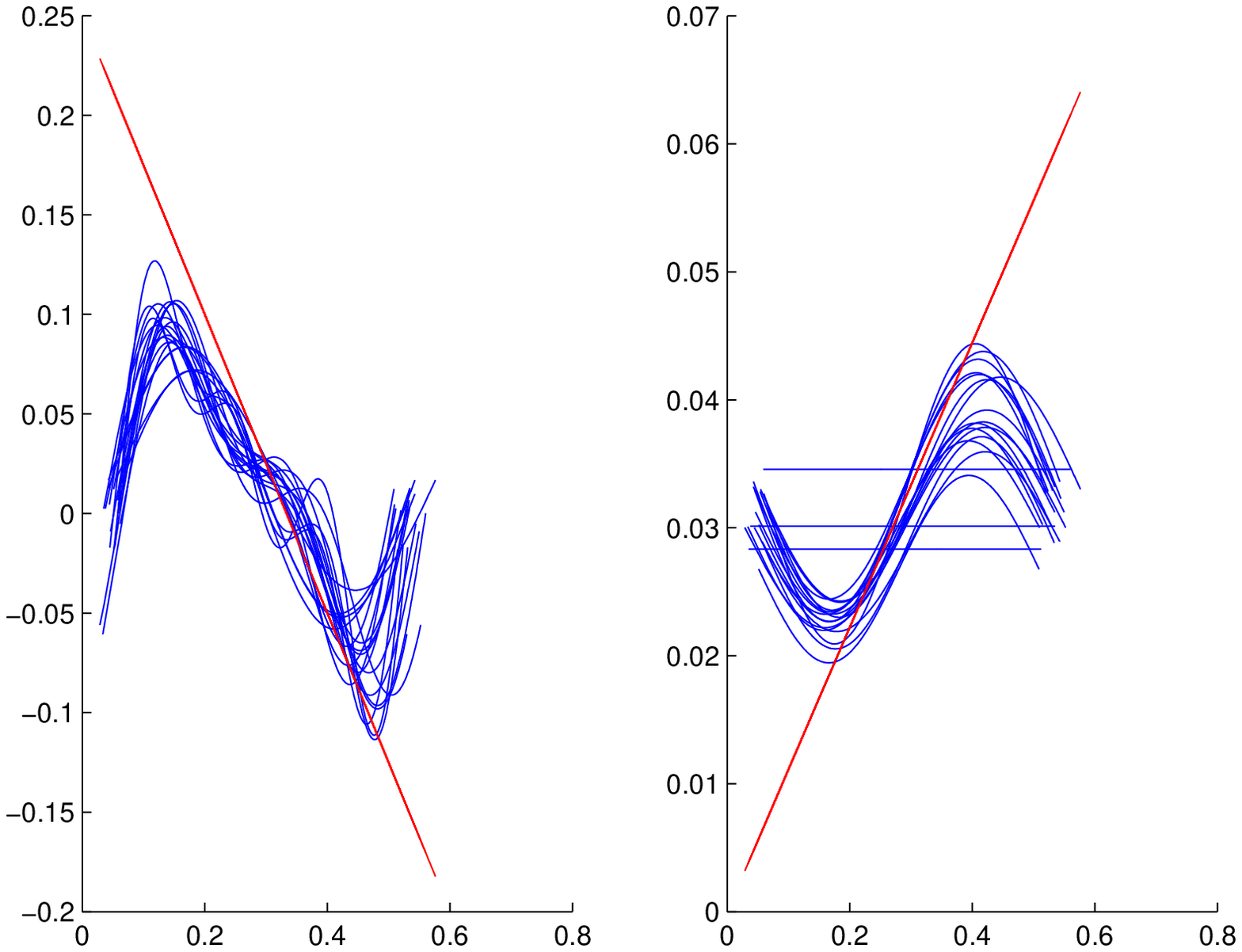}$\\
$\includegraphics[width=1.1 \textwidth,height= 4cm,angle=0]{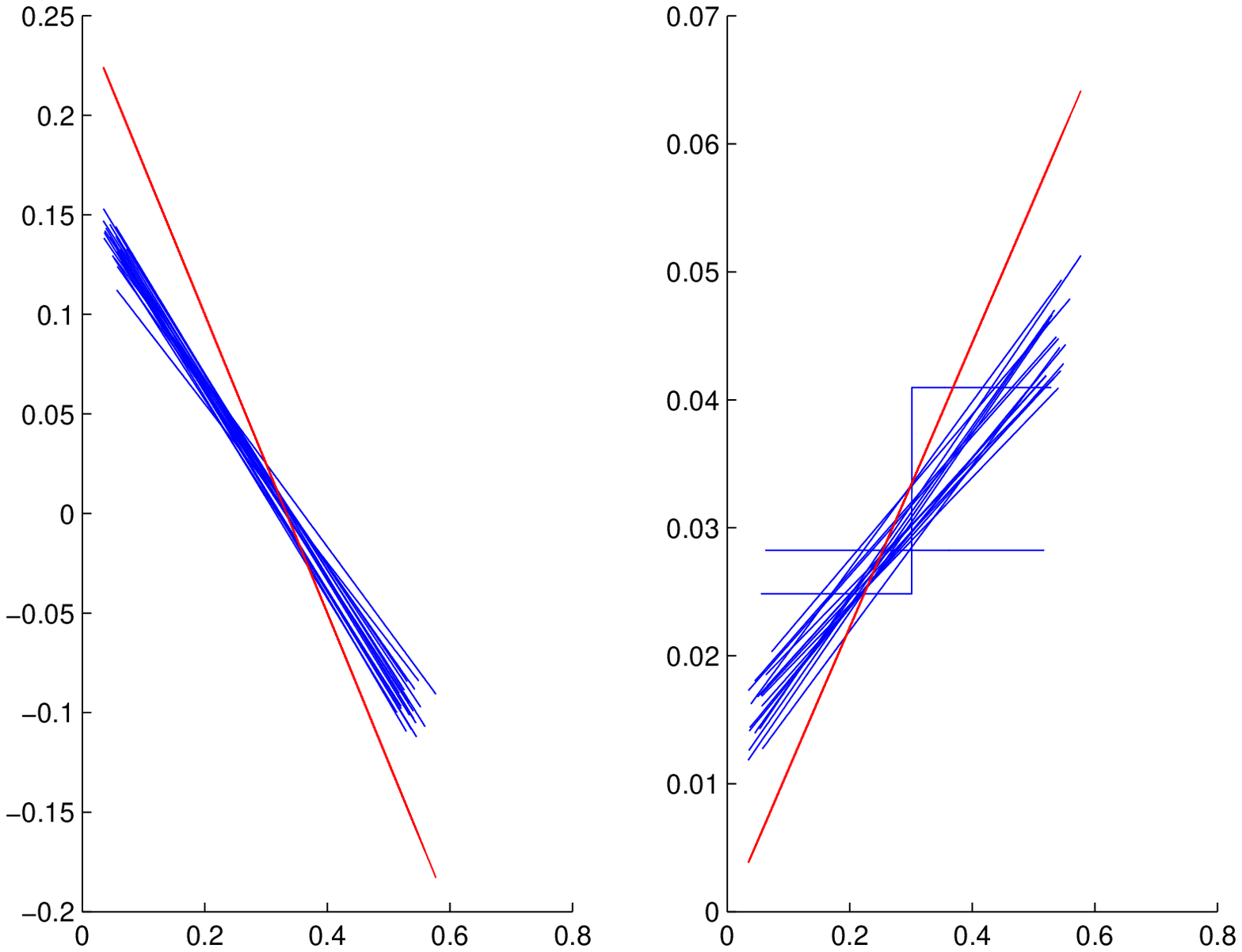}$
\end{tabular}
\caption{Estimation of $b$ (left) and $\sigma^2$ (right) for 20 paths of the CIR process with the trigonometric basis (top) and the piecewise polynomial basis (bottom), $k=250$.}
\end{figure}
The constants $\kappa_1$ and $\kappa_2$ in both drift and diffusion penalties have been set
equal to 2. The term $\hat s_1^2$ replaces $\sigma_1^2/\Delta$ for the estimation of $b$ and $\hat s_2^2$ replaces $\sigma_1^4$ for the estimation of $\sigma^2$. Let us first explain how $\hat s_2^2$ is obtained. We run once the estimation algorithm of $\sigma^2$ with the basis [T] and with a preliminary penalty where $\hat s_2^2$ is taken equal to $2\max_m(\gamma_n^{(2)}(\hat\sigma^2_m))$. This gives a preliminary estimator $\tilde\sigma^2_0$. Afterwards, we take $\hat s_2$ equal to twice the 99.5\%-quantile of $\tilde\sigma^2_0$.
The use of the quantile is here to avoid extreme values.
We get $\tilde\sigma^2$. We use this estimate and 
set $\hat s_1^2=\max_{0\leq k\leq N-1}(\tilde\sigma^2(\hat{\bar V}_k))/\Delta$ for the penalty of $b$. 
\begin{figure}[!ht]\label{fig2} 
$$\includegraphics[width=1.1 \textwidth,height= 4cm,angle=0]{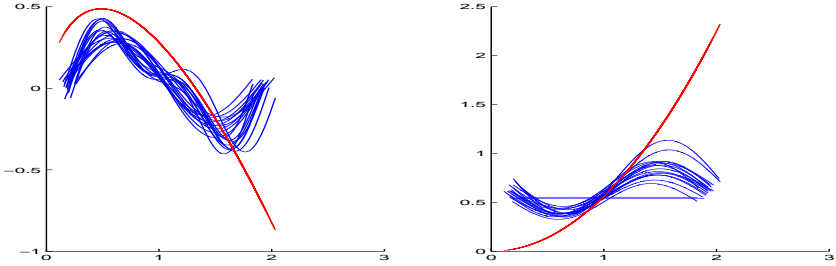}$$
\caption{Estimation of $b$ (left) and $\sigma^2$ (right) for 20 paths of the process $V_t^{(1)}$ (exponential Ornstein Uhlenbeck) with  the trigonometric basis, $k=250$.}
\end{figure}
The results given by our algorithm are described in Figure 1 and 2. We plot in
Figure 1 the true 
function (thick  curve) and 20 estimated  functions (thin curves)  in the case
$b$ and $\sigma^2$ 
when using first the basis [T] and then the basis [GP], in the case of the CIR
process. 
We can see  that the trigonometric basis finds the right  slope in the central
part of the interval, whereas the basis  [GP] in general selects only one bin  and a straight curve,
but with a slightly too small slope. The same type of result holds in Figure 2 for the exponential Orsntein Uhlenbeck process. For comparison with direct or integrated observations of $V$, we refer to Comte {\it et al.}~(2006,2007). It is not surprising that in the case of a stochastic volatility model, empirical results are less satisfactory and require a large number of observations.

We also  give in Tables  1 and 2  results of Monte-Carlo type  experiments. In
Table 1, we show 
the results of the estimation procedure with the basis [T] and the CIR process
when choosing different values of $k$ for  building the quadratic variation. Clearly, there  is an optimal value. If $k$ is too
large, there are 
not  enough observations  left for  the estimation  algorithm. If  $k$  is too
small, bias phenomena appear, 
related  to the  violation  of the  theoretical  assumptions (mainly  $1/k\leq
\Delta$). 
We  repeated the  experiment for  the other  processes and  obtained analogous
results. 
In general, for this sample size, the choice $k=250$ seems to be relevant.
In Table 2, we can see from the  last two columns that the basis [GP] seems to
be better than [T], 
at least  for the CIR process.  The errors are  computed as the mean  over 100
simulated 
paths of the empirical errors (e.g. $(1/N\sum_{i=0}^{N-1} [b(\hat{\bar V}_i)-\tilde b(\hat{\bar V}_i)]^2$ for $b$).

\section{Discussion on the assumptions and proofs}\label{proofs}

\subsection{Proof of Proposition \protect \ref{beta}}
We start with some preliminaries. Let $I_t=\int_0^t V_sds$. The joint process $(V_t, I_t)_{t\geq 0}$ is a two dimensional diffusion satisfying:
$$\left\{ \begin{array}{l} dV_t=b(V_t)dt +\sigma(V_t)dW_t, \; V_0=\eta, \\ dI_t= V_tdt, \;\; I_0=0\end{array}\right.$$
Under regularity assumptions on $b$ and $\sigma$, this process admits a transition density, say $q_t(v_0,i_0,; v,i)$ for the conditional density of $(V_t, I_t)$ given $V_0=v_0$, $I_0=i_0$. This density is w.r.t. the Lebesgue measure on $(0,+\infty)^2$ (see Rogers and Williams~(2000)). We assume that these assumptions hold.

Now, let us set \begin{equation}\label{jell} J_{\ell\delta}= \int_{(\ell -1)\delta}^{\ell \delta}V_sds, \; \ell\geq 1. 
\end{equation} The discrete time process $(V_{\ell\delta}, J_{\ell \delta})_{\ell\geq 1}$  is strictly stationary and Markov. Its one step transition operator is given by the density: $$(v, j)\rightarrow q_{\delta}(v_0,0; v,j):= q_{\delta}(v_0;v,j).$$
Its stationary density is given by $\int \pi(v_0) dv_0 q_{\delta}(v_0;v,j):=\pi_{\delta}(v,j)$. 

Let us set, for $\ell\geq 1$, \begin{equation}\label{zell} Z_{\ell}= X_{\ell\delta}-X_{(\ell -1)\delta}\end{equation} and define $\varepsilon_{\ell}$ by the relation: $Z_{\ell}=J_{\ell \delta}^{1/2}\varepsilon_{\ell}$. Conditionally on $(V_t)_{t\geq 0}$, the random variables (r.v.) $Z_{\ell}, \ell\geq 1$ are independent and $Z_{\ell}$  has distribution ${\mathcal N}(0, J_{\ell\delta})$. Consequently, the r.v $(\varepsilon_{\ell}, \ell \geq 1)$ are i.i.d. with distribution ${\mathcal N}(0,1)$ and the sequence $(\varepsilon_{\ell}, \ell \geq 1)$ is independent of $(V_t)_{t\geq 0}$. Hence $(Z_{\ell})_{\ell\geq 1}$ and $(\hat {\bar V}_i)_{i\geq 0}$ are strictly stationary processes. 
 
\noindent From the 
preliminaries and the above remarks, we deduce that the process $(V_{\ell\delta}, J_{\ell \delta}, \varepsilon_{\ell})_{\ell\geq 1}$ is stationary Markov. Its $\ell$-step transition operator is given by:
 $$Q_{\ell}^{\delta}(v_0;  dv,  dj,  du)= q_{\delta}^{(\ell)}(v_0;  v,j)  n(u)
 dvdjdu$$ 
where $q_{\delta}^{(\ell)}(v_0; v,j)$ is the $\ell$-step transition density 
of  $(V_{\ell\delta}, J_{\ell \delta})$  and $n(u)$  is the  standard gaussian
density. 
The    stationary    density    of    $(V_{\ell\delta},    J_{\ell    \delta},
\varepsilon_{\ell})_{\ell\geq 1}$ is 
$\pi_{\delta}(v,j)n(u)$. Hence 
\begin{eqnarray*} \| Q_{\delta}^{(\ell)}(v_0; dv, dj, du)-\pi_{\delta}(v,j)n(u)dvdjdu\|_{TV} &=& \int |q_{\delta}^{(\ell)}(v_0, v_j) - \pi_{\delta}(v,j))| n(u) dvdjdu\\ &=& 
 \int |q_{\delta}^{(\ell)}(v_0; v,j) - \pi_{\delta}(v,j))|  dvdj.\end{eqnarray*}
 We  may now  use  the  representation of  the  $\beta$-mixing coefficient  of
 strictly stationary Markov processes (see e.g. Genon-Catalot {\it et al.}~(2000)) to compute 
\begin{eqnarray*} \beta_{V_{.\delta}, J_{.\delta}, \varepsilon}(\ell) &=& \int \pi_{\delta}(v_0,j_0)n(u_0) du_0dv_0dj_0 \| Q_{\delta}^{(\ell)}(v_0; dv, dj, du)-\pi_{\delta}(v,j)n(u)dvdjdu\|_{TV}\\ &=& \beta_{V_{.\delta}, J_{.\delta}}(\ell).
\end{eqnarray*}
 Now, we have $\beta_Z(\ell)\leq  \beta_{V_{.\delta}, J_{.\delta}, \varepsilon}(\ell)=\beta_{V_{.\delta}, J_{.\delta}}(\ell)\leq \beta_V((\ell-1)\delta)$. Finally,  $$\beta_{\hat{\bar V}}(i)\leq \beta_Z(ik)\leq \beta_V((ik-1)\delta)\leq c\beta_V(i\Delta). \;\;\; \Box$$

\subsection{Discussion on the assumptions}\label{discu}

Actually, Assumption [A3] is too strong. We only need the existence of moments up to a certain order. 
Let us now discuss [A5]. 
Using the representation $$\hat{\bar V}_0=\frac 1{k\delta}\sum_{\ell=1}^{k}J_{\ell\delta} \; \varepsilon_{\ell}^2,$$ we see that $\hat{\bar V}_0$ has a conditional density given $(V_t, t\geq 0)$. Integrating this density w.r.t. the distribution of $(J_{\ell\delta}, \ell=1, \dots, k)$, we get that $\hat{\bar V}_0$ has a density $\pi^*$. However the formula for $\pi^*$ is untractable.

On the other hand, we can obtain (\ref{bornepi}) by another approach. We have 
$$t^2(\bar V_0)=t^2(V_0)+ (\bar V_0-V_0) (t^2)'(V_0) + \frac 12 (\bar V_0-V_0)^2 \int_0^1 (t^2)"(V_0+u(\bar V_0-V_0))du.$$  Now we use that, for any $t\in S_m$, there exists some constant $C$ such that
 $$\|(t^2)'\|_{\infty} \leq CD_m^2\|t\|^2 \mbox{ and }  \|(t^2)"\|_{\infty}\leq CD_m^3\|t\|^2.$$
 Noting that $|{\mathbb E}\left( \bar V_0-V_0|{\mathcal F}_0\right)|=O(\Delta)$, we get  $| {\mathbb E}[(\bar V_0-V_0)(t^2)'(V_0)]| \leq CD_m^2\Delta\|t\|^2 = O(D_m^2\Delta)$. On the other hand, \begin{eqnarray*} \left|{\mathbb E}\left[(\bar V_0-V_0)^2 \int_0^1 (t^2)"(V_0+u(\bar V_0-V_0))du\right]
 \right|&\leq & \|(t^2)"\|_{\infty} {\mathbb E}[(\bar V_0-V_0)^2] \\ &\leq & CD_m^3\Delta\|t\|^2.
 \end{eqnarray*}
It follows that $|{\mathbb E}(t^2(\bar V_0)-t^2(V_0))|\leq C\Delta D_m^3 \|t\|^2$. Next, 
\begin{eqnarray*} t^2(\hat{\bar V}_0)&=& t^2(\bar V_0)+ (\hat{\bar V}_0-\bar V_0) (t^2)'(V_0) +(\hat{\bar V}_0-\bar V_0) [(t^2)'(\bar V_0) -(t^2)'(V_0)]  \\ && + \frac 12 (\hat{\bar V}_0-\bar V_0)^2 \int_0^1 (t^2)"(\bar V_0+u(\hat{\bar V}_0-\bar V_0))du.\end{eqnarray*} 
By Gloter's~(2007) Proposition 3.1, we have $|{\mathbb E}[(\hat{\bar V}_0-\bar V_0)|V_0]|\leq c\delta (1+V_0)^c$ and 
${\mathbb E}[|\hat{\bar V}_0-\bar V_0|^2]\leq c/k$. 
Hence $$|{\mathbb E}(t^2(\hat{\bar V}_0)-t^2(\bar V_0))|\leq C\|t\|^2(\Delta D_m^2 + \frac{\sqrt{\Delta}D_m^3}{\sqrt{k}} + \frac{D_m^3}{k}).$$ Since $1/k\leq \Delta$
$$|{\mathbb E}(t^2(\hat{\bar V}_0)-t^2(V_0))|\leq C\|t\|^2\Delta D_m^3.$$
As there exist two positive constants $\pi_0$, $\pi_1$ such that $\forall v\in A$, $\pi_0\leq \pi(v)\leq \pi_1$, we obtain $$(\pi_0-C\Delta {\mathcal D}_n^3)\|t\|^2 \leq \|t\|^2_{\pi^*} \leq (\pi_1+C\Delta {\mathcal D}_n^3)\|t\|^2.$$
Under   the   constraint    that   $\Delta{\mathcal   D}_n^3=o(1)$,   we   get
(\ref{bornepi}) for $n$ large enough. 
This constraint is compatible with the other ones, see the discussion after Theorem \ref{maindrift}.


\subsection{Definition of the residuals and their properties}
We have $$R^{(1)}(i+1)=b(\bar V_i)-b(\hat{\bar V}_i)+
R^{(1)}_*((i+1)\Delta)$$ where $R^{(1)}_*$ is the residual term
for $b$ studied in Comte {\it et al.}~(2006, Proposition 3.1) and defined by
$$R^{(1)}_*((i+1)\Delta)= b(V_{(i+1)\Delta})-b(\bar V_i)+ \frac 1{\Delta^2}\int_{(i+1)\Delta}^{(i+3)\Delta} \psi_{(i+1)\Delta} (s) (b(V_s)-b(V_{(i+1)\Delta}))ds.$$
On the other hand, $$R^{(2)}(i+1)=\frac 32\frac{(u_{i+1,k}-u_{i,k})^2}{\Delta} + [\sigma^2(V_{(i+1)\Delta}-\sigma^2(\hat{\bar V}_i)]  +R^{(2)}_*((i+1)\Delta)$$
where $R_*^{(2)}$ is the residual term for $\sigma^2$ studied in Comte {\it et al.}~(2006, Propositions 4.1, 4.2 and 4.3) defined by
$R^{(2)}_*=\sum_{m=1}^3 R_*^{(2,m)}$ with
\begin{eqnarray*}
R_*^{(2,1)}(i\Delta) &=& \frac 3{2\Delta^3} \left(\int_{i\Delta}^{(i+2)\Delta} \psi_{i\Delta}(s)b(V_s)ds\right)^2\\
R_*^{(2,2)}(i\Delta) &=&  \frac 3{\Delta^3} \left(\int_{i\Delta}^{(i+2)\Delta} \psi_{i\Delta}(u)(b(V_u)-b(V_{i\Delta}))du\right) \left( \int_{i\Delta}^{(i+2)\Delta}\psi_{i\Delta}(u)\sigma(V_u)dW_u\right)\\
R_*^{(2,3)}(i\Delta) &=& \frac 3{2\Delta^3}
\int_{i\Delta}^{(i+2)\Delta} \left(\int_s^{(i+2)\Delta}
\psi^2_{i\Delta}(u)du\right)\tau_{b,\sigma}(V_s)ds,
\end{eqnarray*} where $\tau_{b,\sigma}=(\sigma^2/2)(\sigma^2)"+
b(\sigma^2)'$. This decomposition is
obtained by applying Ito's formula and Fubini's theorem. \\
We may now summarize the following useful results, proved in Comte {\it et
al.}~(2006, Propositions 3.1, 4.1, 4.2 and 4.3):
\begin{lem}\label{delta2}
Under Assumptions [A1]-[A2]-[A3], \begin{enumerate} \item For
$\ell=1,2$, for $m=1,2$, for all $i$, ${\mathbb E}\{
[R_*^{(\ell)}(i\Delta)]^{2m}\} \leq c\Delta^{2m\ell}$ where  $c$
is a constant. \item Let $Z^{(1)}_*(i)=(1/\Delta^2)
\int_{i\Delta}^{(i+2)\Delta} \psi_{i\Delta}(s)\sigma(V_s)dW_s$.
For all $i$, ${\mathbb E}([Z^{(1)}_*(i)]^2) \leq
(2/3\Delta){\mathbb E}(\sigma^2(V_0))$. \item For all $i$,
${\mathbb E}([Z^{(2,1)}_i]^2) \leq c_1{\mathbb E}(\sigma^4(V_0))$
and  ${\mathbb E}([Z^{(2,2)}_i]^2) \leq c_2\sigma^2_1\Delta$.
\end{enumerate}
\end{lem}

We also need the following result:
\begin{lem}\label{chapbar} Under assumptions [A1]-[A3], for any integer $i$,
${\mathbb E}[(\bar V_i -\hat{\bar V}_i)^2]={\mathbb E}(u_{i,k}^2)\leq 2{\mathbb E}(V_0^2)/k$ and
${\mathbb E}[(\bar V_i-\hat{\bar V}_i)^4]= {\mathbb E}(u_{i, k}^4)
\leq 56 {\mathbb E}(V_0^4)/k^2$.
\end{lem}
\noindent {\bf Proof of Lemma \ref{chapbar}.} This follows from Proposition 3.1 p.504 in Gloter~(2007).$\Box$

\subsection{Proof of Propositions \ref{propb} and \ref{bornesigma}}
For sake of brevity, we give both proofs at the same time. The
main difference lies in the orders of the expectations and in the
appearance of a specific term in the study of the estimator of
$\sigma^2$. Let us thus define $R_{**}^{(\ell)}$ for $\ell=1,2$ as
$R_{**}^{(1)}=R^{(1)}$ and
$$R_{**}^{(2)}(i+1)=R^{(2)}(i+1)-[\sigma^2(V_{(i+1)\Delta}-\sigma^2(\hat{\bar
V}_i)].$$ Moreover let $T_N^{(1)}(t)=0$ and
$$T_N^{(2)}(t)=\frac 1N \sum_{i=0}^{N-1} (\sigma^2(V_{(i+1)\Delta} -
\sigma^2(\hat{\bar V}_i)) t(\hat{\bar V}_i).$$

Let us consider the set
\begin{equation}\label{omegan}
\Omega_N=\left\{ \omega/
\left|\frac{\|t\|_N^2}{\|t\|^2_{\pi^*}}-1\right|\leq \frac 12,
\;\; \forall t\in \cup_{m,m'\in {\mathcal
M}_n}(S_m+S_{m'})/\{0\}\right\}.\end{equation}  On $\Omega_N$, $\|t\|_{\pi^*}\leq \sqrt{2}\|t\|_N$. From (\ref{deccontrast}), we deduce
\begin{eqnarray*} \|\hat f_m^{(\ell)}-f_A^{(\ell)}\|_N^2
&\leq & \|f_m^{(\ell)}- f_A^{(\ell)}\|_N^2 + \frac
18\|\hat f_m^{(\ell)}-f_m^{(\ell)}\|_{\pi^*}^2 + 16\sup_{t\in S_m,
\|t\|_{\pi^*}=1} [\nu_N^{(\ell)}]^2(t)\\ && +16\sup_{t\in S_m,
\|t\|_{\pi^*}=1}[T_N^{(\ell)}(t)]^2 \\ && + \frac 18\|
\hat f_m^{(\ell)}-f_m^{(\ell)}\|_N^2 + \frac 8N \sum_{i=0}^{N-1}
[R^{(\ell)}_{**}(i+1)]^2
\\ &\leq &
\|f_m^{(\ell)}-f_A^{(\ell)}\|_N^2 + \frac
38\|\hat f^{(\ell)}_m-f^{(\ell)}_m\|^2_N + 16\sup_{t\in S_m,
\|t\|_{\pi^*}= 1} [\nu_N^{(\ell)}]^2(t) \\ &&
+\frac{16}{\pi_0^*}\sup_{t\in S_m, \|t\|=1} [T_N^{(\ell)}(t)]^2 + \frac
8N \sum_{i=0}^{N-1} [R^{(\ell)}_{**}(i+1)]^2.
\end{eqnarray*}
In the last line above, we use the lower bound $\pi_0^*$ introduced in [A5].\\
Setting $B_m(0,1)=\{t\in S_m, \|t\|=1\}$ and
$B_m^{\pi^*}(0,1)=\{t\in S_m, \|t\|_{\pi^*}=1\}$, the following holds on the set $\Omega_N$:
$$\frac 14 \|\hat f^{(\ell)}_m-f^{(\ell)}_A\|_N^2\leq \frac 74\|f^{(\ell)}_m-f^{(\ell)}_A\|_N^2
+ 16\sup_{t\in B_m^{\pi^*}(0,1)} [\nu_N^{(\ell)}]^2(t)+
\frac{16}{\pi_0^*} \sup_{t\in B_m(0,1)} [T_N^{(\ell)}(t)]^2 +
\frac 8N \sum_{i=0}^{N-1} [R^{(\ell)}_{**}(i+1)]^2. $$ We have
the following result:
\begin{lem}\label{lessupnu} Under assumptions  [A1]-[A3] and [A5], if $1/k\leq
  \Delta$, we have, 
for $\ell=1, 2$
$${\mathbb                 E}\left(\sup_{t\in                B_m^{\pi^*}(0,1)}
  [\nu_N^{(\ell)}]^2(t)\right)\leq K\frac{C_{\ell} D_m}{N\Delta^{2-\ell}},$$
with $C_{\ell}={\mathbb E}(\sigma^{2\ell}(V_0))$.
\end{lem}

The Lipschitz condition on $b$ and Lemma
\ref{chapbar} imply that
$${\mathbb E}[(b(\bar V_i)-b(\hat{\bar V}_i))^2]\leq c_b{\mathbb E}[(\bar V_i-\hat{\bar V}_i)^2]\leq 2c_b{\mathbb E}(V_0^2)/k.$$
Consequently, there exists a constant $c$ such that
$${\mathbb E}\left(\frac{8}{N}\sum_{i=0}^{N-1} [R_{**}^{(1)}(i+1)]^2 \right) \leq c(\Delta + k^{-1}). $$
Thus 
$$
{\mathbb E}(\|\hat b_m-b_A\|_N^2\1_{\Omega_N}) \leq  7
\|b_m-b\|_{\pi^*}^2+ \frac{32}{\pi_0^*} {\mathbb E}\left(
\sup_{t\in S_m,\|t\|=1}[\nu_N^{(1)}(t)]^2 \right) + c"(\Delta+
k^{-1}).
$$
By gathering all bounds, we find
$${\mathbb E}(\|\hat b_m-b\|_N^2\1_{\Omega_N}) \leq  7\|b_m-b\|^2_{\pi^*} +
 K\frac{{\mathbb E}(\sigma^2(V_0)) D_m}{N\Delta}(1+\frac 1{k\Delta})  + K'(\Delta+ k^{-1}) .$$

On the other hand, Lemma \ref{delta2} and Lemma \ref{chapbar}
imply that
\begin{eqnarray*} {\mathbb E}(\frac 1N\sum_{i=0}^{N-1} [
R_{**}^{(2)}(i+1)]^2 &\leq& 2{\mathbb E}\left[\frac 1N\sum_{i=0}^{N-1}
\left( [R^{(2)}_*(i+1)]^2 + \frac
94\frac{(u_{i+1,k}-u_{i,k})^4}{\Delta^2}\right) \right] \\ &\leq &
2c\Delta^2 + \frac{36}{\Delta^2} {\mathbb E}(u_{1,k}^4) \leq
C(\Delta^2+\frac 1{k^2\Delta^2}).
\end{eqnarray*}

Next we need to bound ${\mathbb E}\left(\sup_{t\in S_m, \|t\|=1}
[T_N^{(2)}(t)]^2\right)$. This is obtained in the following Lemma:
\begin{lem}\label{tnstar}
Under the Assumptions of Proposition \ref{bornesigma} and if
$1/k\leq \Delta$,  there exists a constant $C$ such that
$${\mathbb E}\left(\sup_{t\in S_m, \|t\|=1} [T_N^{(2)}(t)]^2\right)
\leq C( D_m^2\Delta^2+ D_m^5\Delta^3+ D_m^3/k^2+D_m/(Nk)).$$
\end{lem}

\noindent We can use Lemma 6.1 in Comte {\it et al.}~(2005) to
obtain that, if ${\mathcal D}_n\leq C\sqrt{N\Delta}/\ln(N)$, then
$${\mathbb P}(\Omega_N^c) \leq \frac c{N^4}.$$ This enables to
check that ${\mathbb E}(\|\hat
f_m^{(\ell)}-f^{(\ell)}\|_N^2\1_{\Omega_n^c}) \leq  c/N$ using the  same lines
as the analogous proof given 
p.532 in Comte {\it et al.}~(2007). For this reason, details are omitted. $\Box$

\subsection{Proof of Lemma \ref{lessupnu}.}
\noindent Case $\ell=1$.
Next, let us define ${\mathcal F}_t=\sigma((W_s, B_s), 0\leq
s\leq t, \eta).$ We can use martingale properties to see that, $\forall
t\in S_m$,  $${\mathbb E}(t(\hat{\bar V}_i)Z_{i+1}^{(1)}) =
{\mathbb E}({\mathbb E}(t(\hat{\bar
V}_i)Z_{i+1}^{(1)}|{\mathcal F}_{(i+1)\Delta})) = {\mathbb
E}(t(\hat{\bar V}_i){\mathbb E}(Z_{i+1}^{(1)}|{\mathcal
F}_{(i+1)\Delta}))=0$$ because the last conditional expectation is
zero. Moreover, the same tool shows that the covariance term
${\mathbb E}( t(\hat{\bar V}_i)t(\hat{\bar
V}_{\ell})Z_{i+1}^{(1)}Z_{\ell+1}^{(1)})$ for $\ell \geq
i+2$ is also null by inserting a conditional expectation given
${\mathcal F}_{(\ell +1)\Delta}$. Consequently, it is now easy to
see that \begin{eqnarray*} {\mathbb E}\left( \sup_{t\in
S_m,\|t\|=1}[\nu_N^{(1)}(t)]^2 \right) &\leq & \sum_{j=1}^{D_m} {\mathbb E}[\nu_N^2(\varphi_j)]
\leq \sum_{j=1}^{D_m} {\rm Var}\left[\frac 1{N}\sum_{i=0}^{N-1} \varphi_j(\hat{\bar V}_i) Z^{(1)}_{i+1} \right] \\
&\leq & \frac 2N \sum_{j=1}^{D_m} {\rm
Var}\left(\varphi_{j}(\hat{\bar V}_1)Z^{(1)}_{2}\right)
\\
&\leq & \frac 2N \sum_{j=1}^{D_m} {\mathbb
E}(\varphi^2_{j}(\hat{\bar V}_1) Z^{(1)}_{2})^2) \leq
\frac{2 D_m {\mathbb E}[(Z^{(1)}_{2})^2]}N.
\end{eqnarray*}
Now, Lemma \ref{chapbar} implies that ${\mathbb
E}[(u_{i+2,k}-u_{i+1,k})^2/\Delta^2 = {\mathbb
E}[(u_{i+2,k}^2+u_{i+1,k}^2)/\Delta^2\leq c/(k\Delta^2)$. Then, applying also
Lemma \ref{delta2} $(ii)$, it follows that, with  $${\mathbb E}\left( \sup_{t\in
S_m,\|t\|=1}[\nu_N^{(1)}(t)]^2 \right)\leq
K\frac{D_m}{N\Delta}\left(1+ \frac 1{k\Delta}\right).$$

\noindent Case $\ell=2$.
Next, for the martingale terms, we write
\begin{eqnarray*}
{\mathbb E}(\sup_{t\in B_m^{\pi^*}(0,1)} [\nu_N^{(2)}(t)]^2) &\leq &
\frac 1{\pi_0^*} {\mathbb E}(\sup_{t\in B_m(0,1)}
[\nu_N^{(2)}(t)]^2) \leq \frac 1{\pi_0^*}\sum_{j=1}^{D_m}
{\mathbb E}([\nu_n^{(2)}(\varphi_j)]^2 )\\ &=& \frac
1{\pi_0^*}\sum_{j=1}^{D_m} {\mathbb E}\left(\frac
1N\sum_{i=0}^{N-1} \varphi_{j}(\hat{\bar V}_i) Z^{(2)}_{i+1}\right)^2 \\
&\leq &  \frac 2{\pi_0^*}\sum_{j=1}^{D_m} {\mathbb
E}\left[\left(\frac 1N\sum_{i=0}^{N-1} \varphi_{j}(\hat{\bar
V}_i) (Z_{i+1}^{(2,1)}+Z_{i+1}^{(2,2)})\right)^2\right. \\ &&
\left.\hspace{2cm}  + \left(\frac 9{N\Delta}\sum_{i=0}^{N-1}
\varphi_{j}(\hat{\bar V}_i)(\bar V_{i+2}-\bar
V_i)(u_{i+2,k}-u_{i+1,k}) \right)^2 \right]
\end{eqnarray*}
Both terms are bounded separately. For the first one, we use that, for $r=1,2$
$${\rm cov}(\varphi_{j}(\hat{\bar V}_i)
Z^{(2,r)}_{i+1}, \varphi_{j}(\hat{\bar V}_{\ell})
Z^{(2,r)}_{\ell+1})=0$$ if $\ell\geq i+2$, by inserting a
conditional expectation with respect to ${\mathcal
F}_{(\ell+1)\Delta}$. Now, for $r=1,2$,
\begin{eqnarray*}
&& \sum_{j=1}^{D_m} {\mathbb E}\left[\left(\frac
1N\sum_{i=0}^{N-1} \varphi_{j}(\hat{\bar V}_i)
Z^{(2,r)}_{i+1}\right)^2\right]  \leq \frac  1{N^2}  \sum_{j=1}^{D_m} {\mathbb
E}\left(\sum_{0\le i,\ell \le N-1}
\varphi_{j}(\hat{\bar V}_{i})
Z^{(2,r)}_{i+1}\varphi_{j}(\hat{\bar V}_{\ell})
Z^{(2,r)}_{\ell+1}\right)
\\
&=& \frac 1{N^2}\sum_{j=1}{D_m}{\mathbb
E}\left\{\sum_{i=0}^{N-1}\left[ \varphi_{j}^2(\hat{\bar V}_i)[Z^{(2,r)}_{i+1}]^2+
  \varphi_{j}(\hat{\bar V}_i)Z^{(2,r)}_{i+1} \varphi_{j}(\hat{\bar V}_{i+1})
  Z^{(2,r)}_{i+2}\right]\right\}
   \\
  &\leq & \frac 2N \|\sum_{j=1}^{D_m}
  \varphi_{j}^2\|_{\infty} {\mathbb E}[(Z^{(2,r)}_{2})^2]
\leq   2\frac{D_m}{N} [\tilde c_1 {\mathbb
E}(\sigma^4(V_0)) + \tilde c_2\Delta ]
\end{eqnarray*}
by using Lemma \ref{delta2}.

For the second part, let us define the filtration generated by
$B$ and the whole path of $V$, i.e. $${\mathcal
G}^V_t=\sigma(V_s, s\in {\mathbb R}^+, B_s, s\leq t) =\sigma(W_s,
s\in {\mathbb R}^+, B_s, s\leq t, \eta).$$ Now we observe that
\begin{eqnarray*} {\mathbb E}(t(\hat{\bar V}_i)(\bar V_{i+2}-\bar
V_{i+1})u_{i+1,k}) &=&
{\mathbb E}\left[{\mathbb E}(t(\hat{\bar V}_i)(\bar V_{i+2}-\bar V_{i+1})u_{i+1,k})|{\mathcal G}^V_{(i+1)\Delta})\right] \\
&=& {\mathbb E}\left[t(\hat{\bar V}_i)(\bar V_{i+2}-\bar V_{i+1}) {\mathbb E}(u_{i+1,k})|{\mathcal G}^V_{(i+1)\Delta})\right] \\
&=& 0\end{eqnarray*} as  ${\mathbb E}(u_{i+1,k})|{\mathcal
G}^V_{(i+1)\Delta})=0$. Moreover for any $\ell >i$,
$${\mathbb E}(t(\hat{\bar V}_i)(\bar V_{i+2}-\bar V_{i+1})u_{i+1,k}t(\hat{\bar V}_{\ell})(\bar V_{\ell+2}-\bar V_{\ell+1})u_{\ell+1,k})) =0$$ by inserting a conditional expectation with respect to ${\mathcal G}^V_{(\ell+1) \Delta}$. The last remark is that one can easilty see that $${\mathbb E}[(\bar V_{i+1}-\bar V_i)^4] \leq \frac 1{\Delta^4} {\mathbb E}\left[\left(\int_{(i+1)\Delta}^{(i+2)\Delta} (V_s-V_{s-\Delta})ds\right)^4 \right] \leq C\Delta^2.$$
Now we have
\begin{eqnarray*}
\sum_{j=1}^{D_m} {\mathbb E} \left(\frac
1{N\Delta}\sum_{i=0}^{N-1} \varphi_{j}(\hat{\bar V}_i)(\bar
V_{i+2}-\bar V_i)u_{i+1,k} \right)^2 &=& \frac 1{N^2\Delta^2}
\sum_{j=1}^{D_m} \sum_{i=0}^{N-1} {\mathbb E} \left(
\varphi_{j}^2(\hat{\bar V}_i)(\bar V_{i+2}-\bar
V_i)^2u_{i+1,k}^2 \right) \\ &\leq & \frac{D_m}{N\Delta^2}
{\mathbb E}^{1/2} [(\bar V_{2}-\bar V_1)^4] {\mathbb
E}^{1/2}[u_{2,k}^4]\\ & \leq &  C\frac{D_m}{N}\frac 1{k\Delta}.
\end{eqnarray*}
The second part of this term  can be treated in the same way, and it follows that if $1/k\leq \Delta$, then this term is less than $C'D_m/N$. $\Box$

\subsection{Proof of Lemma \ref{tnstar}.} Let us recall that we know
from Comte {\it et al.}~(2006) that
$$T_N^*(t)=\frac 1N \sum_{i=0}^{N-1} (\sigma^2(V_{(i+1)\Delta} -
\sigma^2(\bar V_i)) t(\bar V_i)$$ is such that $${\mathbb
E}(\sup_{t\in B_m(0,1)} [T_N^*(t)]^2) \leq C(D_m^2\Delta^2 +
D_m^5\Delta^3).$$ Here, we write that $T_N^{(2)}(t) = T_N^{(2,1)}(t) +
T_N^{(2,2)}(t) +  T_N^{(2,3)}(t)+ T_N^*(t)$ with
$$T_N^{(2,1)}(t)=\frac 1N\sum_{i=0}^{N-1} [t(\hat{\bar V}_i)-t(\bar V_i)][\sigma^2(\hat{\bar V}_i) -\sigma^2(\bar V_i)], \;\;
T_N^{(2,2)}(t) = \frac 1N\sum_{i=0}^{N-1} t(\bar V_i)[\sigma^2(\hat{\bar
V}_i) -\sigma^2(\bar V_i)],$$
$$T_N^{(2,3)}(t) = \frac 1N \sum_{i=0}^{N-1} [t(\hat{\bar V}_i)-t(\bar V_i)][\sigma^2(\bar V_i) -\sigma^2(V_{(i+1)\Delta})].$$
We shall use the following decompositions obtained by the Taylor
formula:
$$\sigma^2(\hat{\bar V}_i) - \sigma^2(\bar V_i)= (\hat{\bar V}_i-\bar V_i)(\sigma^2)'(\bar V_i) + R_i,
t(\hat{\bar V}_i)-t(\bar V_i)= (\hat{\bar V}_i-\bar V_i)t'(\bar
V_i) + S_i(t)$$ with ${\mathbb E}(R_i^2)\leq C/k^2$ and ${\mathbb
E}(R_i^4)\leq C/k^4$ if $(\sigma^2)"$ is bounded, and ${\mathbb
E}\left(\sup_{t\in B_m(0,1)} S_i(t)^2\right) \leq CD_m^5/k^2$,
${\mathbb E}^{1/2}\left(\sup_{t\in B_m(0,1)} S_i(t)^4\right) \leq
CD_m^5/k^2$ because $\|t"\|_{\infty}^2\leq CD_m^5\|t\|^2$. Now,
the three terms can be studied as follows.
First 
\begin{eqnarray*} T_N^{(2,1)}(t)&=& \frac
1N\sum_{i=0}^{N-1}(\hat{\bar V}_i-\bar V_i)^2 (t')(\bar
V_i)(\sigma^2)'(\bar V_i) + \frac 1N\sum_{i=0}^{N-1} (\hat{\bar
V}_i-\bar V_i)t'(\bar V_i)R_i \\ && + \frac
1N\sum_{i=0}^{N-1}(\hat{\bar V}_i-\bar V_i)(\sigma^2)'(\bar V_i)S_i(t)
+ \frac 1N\sum_{i=0}^{N-1} R_iS_i(t) \\ &:=& T_N^{(2,1,1)}(t) +
T_N^{(2,1,2)}(t)+T_N^{(2,1,3)}(t) + T_N^{(2,1,4)}(t),\end{eqnarray*} and
we bound each term successively. Clearly by Schwarz inequality
applied to each term, we find, $${\mathbb E}(\sup_{t\in B_m(0,1)}
[T_N^{(2,1,1)}(t)]^2) \leq C{\mathbb E}^{1/2}(\bar
V_1^4)\frac{D_m^3}{k^2}$$ using that $\|t'\|_{\infty}^2\leq
CD_m^3\|t\|^2$,  $${\mathbb E}(\sup_{t\in B_m(0,1)}
[T_N^{(2,1,2)}(t)]^2) \leq C\frac{D_m^3}{k^3}, \;\;\; {\mathbb
E}(\sup_{t\in B_m(0,1)} [T_N^{(2,1,3)}(t)]^2) \leq C{\mathbb
E}^{1/2}(\bar V_1^4)\frac{D_m^5}{k^3},$$ and
$${\mathbb E}(\sup_{t\in B_m(0,1)} [T_N^{(2,1,4)}(t)]^2) \leq C\frac{D_m^5}{k^4}.$$ Therefore, if $1/k\leq \Delta$,
${\mathbb E}(\sup_{t\in B_m(0,1)} [T_N^{(2,1)}(t)]^2) \leq C
(D_m^3/k^2+ D_m^5/k^3)$.

Next, we write that 
\begin{eqnarray*} T_N^{(2,2)}(t) &=& \frac
1N\sum_{i=0}^{N-1} t(\bar V_i)(\sigma^2)'(\bar V_i)(\hat{\bar
V}_i)-\bar V_i) + \frac 1N\sum_{i=0}^{N-1} t( \bar V_i)R_i \\ &=&
T_N^{(2,2,1)}(t) + T_N^{(2,2,2)}(t).\end{eqnarray*} 
We obtain easily that 
$${\mathbb E}(\sup_{t\in B_m(0,1)} [T_N^{(2,2,2)}(t)]^2)\leq
{\mathbb E}(\sup_{t\in B_m(0,1)}\|t\|_{\infty}^2 \frac
1N\sum_{i=1}^NR_i^2)\leq \Phi_0^2D_m {\mathbb E}(R_1^2)\leq
CD_m/k^2,$$ a term which is negligible with respect to the
previous ones. 

Then $(\hat{\bar V}_i-\bar V_i)\psi(\bar V_i)$ is a martingale increment with respect to the filtration $({\cal G}_t^V)$, for any
measurable function $\psi$. In particular,
\begin{eqnarray*} {\mathbb E}[(\hat{\bar V}_i-\bar V_i)\psi(\bar V_i)] &=& {\mathbb E}[{\mathbb E}[(\hat{\bar V}_i-\bar V_i)\psi(\bar V_i)|{\mathcal G}^V_{i\Delta}]] \\ &=& {\mathbb E}[\psi(\bar V_i) {\mathbb E}[(\hat{\bar V}_i-\bar V_i)|{\mathcal G}^V_{i\Delta}]] =0\end{eqnarray*} since ${\mathbb E}(\hat{\bar V}_i|{\mathcal G}_{i\Delta}^V)=\bar V_i$. In the same way, for $i<\ell$,
$${\mathbb E} \left((\hat{\bar V}_i-\bar V_i)\psi(\bar V_i)(\hat{\bar V}_{\ell}-\bar V_{\ell})\psi(\bar V_{\ell})\right)=0$$ by inserting a conditional expectation with respect to ${\mathcal G}_{\ell \Delta}^V$.
Therefore
\begin{eqnarray*}
{\mathbb E}(\sup_{t\in B_m(0,1)} [T_N^{(2,2,1)}(t)]^2)&\leq
&\sum_{j=1}^{D_m} {\mathbb E}\left(\frac 1N\sum_{i=0}^{N-1}
\varphi_{j}(\bar V_i)(\sigma^2)'(\bar V_i) (\hat{\bar
V}_i-\bar V_i)\right)^2 \\ &=& \sum_{j=1}^{D_m} \frac
1N{\mathbb E}\left(\varphi_{j}(\bar V_1)(\sigma^2)'(\bar
V_1) (\hat{\bar V}_1-\bar V_1\right)^2\\ &\leq & \frac 1N {\mathbb
E}\left((\sum_{j=1}^{D_m}\varphi_{j}^2(\bar V_1))
[(\sigma^2)'(\bar V_1)]^2 (\hat{\bar V}_1-\bar V_1)^2\right) \\
&\leq & \frac{D_m}{N}{\mathbb E}^{1/2}[(\sigma^2)'(\bar
V_1)^4]{\mathbb E}^{1/2}[ u_{1,k}^4] \leq C{\mathbb E}^{1/2}(\bar
V_1^4)\frac{D_m}{Nk}.
\end{eqnarray*}

For the last term, we write
$T_N^{(2,3)}(t)=T_N^{(2,3,1)}(t)+T_N^{(2,3,2)}(t)$ where
$$T_N^{(2,3,1)}(t)=(1/N)\sum_{i=0}^{N-1} (\hat{\bar V}_i-\bar
V_i)t'(\bar V_i)(\sigma^2(\bar V_i)-\sigma^2(V_{(i+1)\Delta})),$$
$$T_N^{(2,3,2)}(t) =(1/N)\sum_{i=0}^{N-1} S_i(t)(\sigma^2(\bar
V_i)-\sigma^2(V_{(i+1)\Delta})).$$  Moreover, we know from Comte
{\it et al.}~(2006) that ${\mathbb E}[(\sigma^2(\bar
V_i)-\sigma^2(V_{(i+1)\Delta}))^2]\leq {\mathbb
E}^{1/2}[(\sigma^2(\bar V_i)-\sigma^2(V_{(i+1)\Delta}))^4]\leq
C\Delta$. Now, for $T^{(2,3,1)}_N(t)$, we proceed as for
$T^{(2,2,1)}_N(t)$ since both have the same martingale property w.r.t.  ${\mathcal G}_{s}^V$. We get
\begin{eqnarray*}{\mathbb E}(\sup_{t\in B_m(0,1)} [T_N^{(2,3,1)}(t)]^2)&\leq & \sum_{j=1}^{D_m} {\mathbb E}\left(\frac 1N\sum_{i=0}^{N-1} \varphi_{j}'(\bar V_i)(\hat{\bar V}_i-\bar V_i)(\sigma^2(\bar V_i)-\sigma^2(V_{(i+1)\Delta}))\right)^2
\\ &\leq & \frac 1N \sum_{j=1}^{D_m} {\mathbb E}\left( (\varphi_{j}')^2(\bar V_1) (\hat{\bar V}_1-\bar V_1)^2
(\sigma^2(\bar V_1)-\sigma^2(V_{2\Delta}))^2\right)\\ &\leq &
\frac{CD_m^3}N {\mathbb E}^{1/2}(u_{1,k}^4){\mathbb E}^{1/2}[
(\sigma^2(\bar V_1)-\sigma^2(V_{2\Delta}))^4]\\ &\leq &
C\frac{D_m^3\Delta}{Nk}
\end{eqnarray*} as $\sum_{j}(\varphi_{j}')^2(x)\leq CD_m^3$.
Using $D_m^2\leq N\Delta$ and $1/k\leq \Delta$ implies ${\mathbb
E}(\sup_{t\in B_m(0,1)} [T_N^{(2,3,1)}(t)]^2)\leq CD_m\Delta^3$.
On the other hand,  ${\mathbb E}(\sup_{t\in B_m(0,1)}
[T_N^{(2,3,2)}(t)]^2)\leq CD_m^5\Delta/k^2\leq CD_m^5\Delta^3$, as
$1/k\leq \Delta$.

By gathering and comparing all terms and assuming that $1/k\leq
\Delta$, we obtain the bound given in Lemma \ref{tnstar}.$\Box$

\subsection{Proof of Theorem \protect \ref{maindrift}}
The proof of this theorem relies on the following Bernstein-type
Inequality:
\begin{lem}\label{bern} Under the assumptions of Theorem \ref{maindrift}, for any positive
numbers $\epsilon$ and $v$, we have $${\mathbb P}\left[
\sum_{i=0}^{N-1} t(\hat{\bar V}_i)Z^{(1)}_{(i+1)\Delta} \geq N\epsilon,
\|t\|_N^2\leq v^2\right] \leq \exp\left(-\frac{N\Delta
\epsilon^2}{2\sigma_1^2v^2}\right).$$
\end{lem}
\noindent {\bf Proof of Lemma \ref{bern}:} Noting that $W$ is a Brownian motion with respect to the augmented filtration ${\mathcal F}_s=\sigma((B_u, W_u), u\leq s, \eta)$, the proof is obtained as the analogous proof in Comte {\it et al.}~(2007), Lemma 2 p.533. $\Box$

Now we turn to the proof of Theorem \ref{maindrift}.\\
As in the proof of Proposition \ref{propb}, we have to split
$\|\tilde b-b_A\|_N^2= \|\tilde b-b_A\|_N^2\1_{\Omega_N} +
\|\tilde b-b_A\|_N^2\1_{\Omega_N^c}$. For the study on
$\Omega_N^c$, the end of the proof of Proposition \ref{propb} can
be used.

Now, we focus on what happens on $\Omega_N$. From the definition
of $\tilde b$, we have, $\forall m\in {\mathcal M}_n$,
$\gamma_N(\hat b_{\hat m}) + {\rm pen}(\hat m)\leq \gamma_N(b_m)+
{\rm pen}(m)$. We proceed as in the proof of Proposition
\ref{propb} with some additional penalty terms and obtain
\begin{eqnarray*}
{\mathbb E}(\|\hat b_{\hat m}-b_A\|_N^2\1_{\Omega_N}) &\leq & 7
\|b_m-b_A\|^2_{\pi^*}+ {\rm pen}(m) +
32 {\mathbb E}\left( \sup_{t\in S_m+S_{\hat m},\|t\|_{\pi^*}=1}[\nu_N^{(1)}(t)]^2 \1_{\Omega_N}\right) \\
&& - {\mathbb E}({\rm pen}(\hat m))  + 32 c'\Delta.\end{eqnarray*}
The difficulty here is to control the supremum of $\nu_N^{(1)}(t)$ on
a random ball (which depends on the random $\hat m$). This is done by setting $\nu_N^{(1)}=\nu_N^{(1,1)} +
\nu_N^{(1,2)}$, with $$\nu_N^{(1,1)}(t)=\frac 1N\sum_{i=0}^{N-1}
Z_{(i+1)\Delta}^{(1)} t(\hat{\bar V}_i), \;\; \nu_N^{(1,2)}(t) = \frac
1N\sum_{i=0}^{N-1} t(\hat{\bar V}_i)
\left(\frac{u_{i+2,k}-u_{i+1,k}}{\Delta}\right).$$
We use the martingale property of $\nu_N^{(1,1)}(t)$ and a
rough bound for $\nu_N^{(1,2)}(t)$ as follows.

For $\nu_N^{(1,2)}$, we simply write, as previously
\begin{eqnarray*} {\mathbb E}\left(\sup_{t\in S_m+S_{\hat m},
\|t\|_{\pi^*}=1}[\nu_n^{(1,2)}(t)]^2\right) &\leq & \frac
1{\pi_0^*}{\mathbb E}\left(\sup_{t\in {\mathcal S}_n,
\|t\|=1}[\nu_n^{(1,2)}(t)]^2\right) \\ &\leq & \frac 1{\pi_0^*}
\sum_{j=1}^{{\mathcal D}_n} {\mathbb
E}[(\nu_N^{(2)}(\varphi_{j}))^2] \\ &\leq &
\frac{4{\mathcal D}_n}{\pi_0^* N} {\mathbb
E}[(u_{1,k}/\Delta)^2]\leq \frac{4{\mathcal D}_n}{\pi_0^*
Nk_n\Delta^2} \leq \frac{4}{\pi_0^*} \frac 1{k_n\Delta}.
\end{eqnarray*} For $\nu_N^{(1,1)}$, let us denote by $$G_m(m')=
\sup_{t\in S_m+S_{m'},\|t\|_{\pi^*}=1}\nu_N^{(1,1)}(t)$$ the
quantity to be studied. Introducing a function $p(m,m')$, we first
write
\begin{eqnarray*}
G_m^2(\hat m)\1_{\Omega_N} &\leq & [(G_m^2(\hat m)-p(m,\hat m))\1_{\Omega_N}]_+ + p(m, \hat m)\\
& \leq &  \sum_{m'\in {\mathcal M}_n}
[(G_m^2(m')-p(m,m'))\1_{\Omega_N}]_+ + p(m, \hat m).
\end{eqnarray*}
Then pen is chosen such that $32 p(m, m')\leq {\rm pen}(m) +{\rm
pen}(m')$. More precisely, the next Proposition determines the
choice of $p(m,m')$ which in turn will fix the penalty.
\begin{prop}\label{dev} Under the assumptions of Theorem \ref{maindrift}, there exists a numerical constant $\kappa_1$ such that, for
$p(m,m')= \kappa_1 \sigma^2_1 (D_m+D_{m'})/(n\Delta)$,  we have
$${\mathbb E}[(G_m^2(m')-p(m,m'))\1_{\Omega_N}]_+\leq c \sigma_1^2
\frac{e^{-D_{m'}}}{N\Delta}.$$
\end{prop}
\noindent {\bf Proof of Proposition \ref{dev}.} The result of
Proposition \ref{dev} follows from the inequality of Lemma
\ref{bern} by the ${\mathbb L}^2$-chaining technique used in
Baraud et al. (2001b) (see Section 7 p.44-47,  Lemma 7.1, with $s^2=\sigma_1^2/\Delta$).~~$\Box$\\

It is easy to see that the result of Theorem \ref{maindrift}
follows from Proposition \ref{dev} with ${\rm pen}(m) =  \kappa
\sigma_1^2 D_m/(N\Delta).$ $\Box$

 \subsection{Proof of Theorem \protect \ref{mainvol}}
The lines of the proof are the same as the ones of Theorem
\ref{maindrift}. Moreover, they follow closely the analogous proof of Theorem 2 p.524 in Comte {\it et al.}~(2007), see also Comte {\it et al.}~(2006). Therefore, we omit it.

\end{document}